\documentclass[a4paper]{article}

\usepackage[margin=1in]{geometry} 

\usepackage{amsmath}
\usepackage{amsthm}
\usepackage{amssymb}

\usepackage[utf8]{inputenc}
\usepackage{hyperref}
\hypersetup{
	unicode,
	pdfauthor={Author One, Author Two, Author Three},
	pdftitle={A simple article template},
	pdfsubject={A simple article template},
	pdfkeywords={article, template, simple},
	pdfproducer={LaTeX},
	pdfcreator={pdflatex}
}

\usepackage[sort&compress,numbers,square]{natbib}
\theoremstyle{plain}
\newtheorem{theorem}{Theorem}

\theoremstyle{definition}
\newtheorem{definition}[theorem]{Definition}

\newtheorem{remark}[theorem]{Remark}

\newtheorem{eg}[equation]{Example}

\usepackage{graphicx, color}
\graphicspath{{fig/}}
\usepackage{mathtools}
\usepackage{lipsum}
\usepackage{tikz}
\usetikzlibrary{cd,positioning}
\usetikzlibrary{calc}
\usetikzlibrary{decorations.pathmorphing}
\tikzset{curve/.style={settings={#1},to path={(\tikztostart)
    .. controls ($(\tikztostart)!\pv{pos}!(\tikztotarget)!\pv{height}!270:(\tikztotarget)$)
    and ($(\tikztostart)!1-\pv{pos}!(\tikztotarget)!\pv{height}!270:(\tikztotarget)$)
    .. (\tikztotarget)\tikztonodes}},
    settings/.code={\tikzset{quiver/.cd,#1}
        \def\pv##1{\pgfkeysvalueof{/tikz/quiver/##1}}},
    quiver/.cd,pos/.initial=0.35,height/.initial=0}

\tikzset{tail reversed/.code={\pgfsetarrowsstart{tikzcd to}}}
\tikzset{2tail/.code={\pgfsetarrowsstart{Implies[reversed]}}}
\tikzset{2tail reversed/.code={\pgfsetarrowsstart{Implies}}}
\tikzset{no body/.style={/tikz/dash pattern=on 0 off 1mm}}

\DeclareMathOperator{\rank}{rank}
\DeclareMathOperator\supp{supp}

\usepackage{mathrsfs} 

\title{Algebraic approaches for the decomposition of reaction networks and the determination of existence and number of steady states}
\author{Joseph M. Sauder$^1$ \and Bruce P. Ayati$^1$ \and Ryan Kinser$^1$\thanks{Author Three was partially supported by National Science Foundation Award No. DMS-2303334.}}

\date{
	$^1$Department of Mathematics, University of Iowa \\%
}

\begin{document}
	\maketitle
	
	\begin{abstract}
		Chemical reaction network theory provides powerful tools for rigorously understanding chemical reactions and the dynamical systems and differential equations that represent them. A frequent issue with mathematical analyses of these networks is the reliance on explicit parameter values which in many cases cannot be determined experimentally. This can make analyzing a dynamical system infeasible, particularly when the size of the system is large. One approach is to analyze subnetworks of the full network and use the results for a full analysis.
		
		Our focus is on the equilibria of reaction networks.  Gröbner basis computation is a useful approach for solving the polynomial equations which correspond to equilibria of a dynamical system. We identify a class of networks for which Gröbner basis computations of subnetworks can be used to reconstruct the more expensive Gröbner basis computation of the whole network. We compliment this result with tools to determine if a steady state can exist, and if so, how many.
		
	\end{abstract}

	
	\section{Introduction}
	\label{sec:intro}
	
	Chemical reaction network theory primarily consists of models of chemical or biochemical phenomena based on ordinary differential equations (ODEs). The kinetics associated with a network determine the rate at which the reactions occur, and once specified they determine the ODE system. Horn and Jackson \cite{Horn_Jackson_1972} showed that a network endowed with mass-action kinetics always induces a polynomial ODE system. The coefficients for these polynomials are rate constants associated with the reactions. A goal of reaction network theory is to provide qualitative results applicable to reaction networks without fixing parameters, so that these results are applicable to networks with a range of parameters.
	
Chemical concentrations cannot always be measured or determined experimentally during the course of a reaction. Taking a measurement may alter the reaction.  As a result, longitudinal data is not always available. In the absence of longitudinal data, parameter estimation must be based on data available at steady state. 
When comparing reaction network models we will examine how well they each fit experimental data at steady states, by looking at the steady-state ideal induced by the reaction network. 
Even at steady state, it may not be possible to measure the concentrations of certain chemical species.  In this case, one approach is to remove those chemicals from the network model. The remaining system can then be parameterized from experimental data.  
	To compensate for the removal of these chemical species, their rate constants are incorporated into the reaction network model by replacing mass-action terms with rational expressions, such as Monod or Michaelis-Menten terms.  The coefficients in the reaction-rate functions would then need to be estimated using measurements of chemical concentrations at different time points, assuming this data is available.
	
	We will use an approach  that first decomposes the network into subnetworks before completely removing a problematic species from a subnetwork \cite{Gross_Harrington_Meshkat_Shiu_2020}.  This decomposition has a biological interpretation and context for the corresponding reaction network and subnetworks.
	
	Beyond advantages with respect to data availability, the steady states of reaction networks are often the subject of an inquiry.  In particular the existence of positive steady states, the number of steady states, and the stability of steady states are often of interest. Even for relatively simple or small polynomial ODE systems these questions can be difficult.

	We also survey and apply some general algebraic methods for answering questions regarding the existence and number of positive steady states. The strength of these tools should be evident from the size and complexity of the particular reaction networks to which we apply them. We showcase this by applying a methodology from \cite{multistationarity_regions2017} to analyze the multistationarity of the n-site phosphorylation dephosphorylation cycle, which is a Post Translational Modification (PTM) networks. We also apply some deficiency theory to analyze the equilibria of a Mass-Action Chemostat model. 
		
	\subsection{Preliminaries}
	\label{sec:pre}
	
	In this section we will formally define a reaction network and its attendant parts. We denote the set of $n$-tuples of positive real numbers by $\mathbb{R}_+^n$, and for $x,y \in \mathbb{R}_+^n$ we write $x^y = (x_1^{y_1}, x_2^{y_2},...,x_n^{y_n})$. We write $\mathcal{S}=\{ X_1, X_2, \dotsc, X_n \}$, and identify $\mathbb{R}^\mathcal{S}:=\mathbb{R}^n$.
Following the conventions used by Feinberg in \cite{Feinberg_2019_book}, we denote a reaction network by $\mathcal{N} = \{ \mathcal{S}, \mathcal{C}, \mathcal{R} \}$ where $\mathcal{S}$ is the set of \textbf{species} in the network;  $\mathcal{C}$ is a set of vectors in $\overline{\mathbb{R}_+^{\mathcal{S}}}$, called the \textbf{complexes} in the network;  $\mathcal{R} \subset \mathcal{C} \times \mathcal{C}$ is a subset called the \textbf{reactions} of the network, where $(y,y) \not\in \mathcal{R}$ for all $y \in \mathcal{C}$, and we write $y \to y'$ when $(y,y') \in \mathcal{R}$. 

For $y \to y'$, the corresponding \textbf{reaction vector} is $y' -  y \in \overline{\mathbb{R}}^{\mathcal{S}}$.
The \textbf{stoichiometric subspace} is $S \coloneqq \text{span}\{y'-y \in \mathbb{R}^{\mathcal{S}}:y \to y' \}$.
A \textbf{kinetics} for a reaction network  is an assignment of a continuously differentiable rate function $\mathcal{K}_{y \to y'}: \overline{\mathbb{R}}_+^{\mathcal{S}} \to \overline{\mathbb{R}}_+$
such that $\mathcal{K}_{y \to y'}(x) > 0 $ if and only if $\supp(y) \subset \supp(x)$.
A reaction network $\mathcal{N}$ together with a kinetics $\mathcal{K}$ is called a \textbf{kinetic system} (KS), and the associated system of ODEs is
\begin{equation}
\dot{x} = \sum_{\mathcal{R}} \mathcal{K}_{y \to y'}(x) (y'-y).
\end{equation}

A kinetics $\mathcal{K}$ for a reaction network is \textbf{mass-action} if, for each $y \to y'$, there exists a  \textbf{rate constant} $k_{y \to y'} \in \mathbb{R}_+$ such that
		\begin{equation} \label{def:mass-action kinetics} 
			\mathcal{K}_{y \to y'}(x) = k_{y \to y'}x^y  \qquad \left(x^y := \prod_{s \in \mathcal{S}} x_s^{y_s}\right),
		\end{equation}
where $y_s$ is the \textbf{stoichiometric coefficient} for species $s$ in $y$, and $x_s$ is the \textbf{molar concentration} of species $s$.
All the reaction networks will be assumed to have mass-action kinetics unless explicitly stated otherwise.
The \textbf{species formation rate function} is 
		\begin{equation} \label{def:species formation function}
			f(x) \coloneqq \sum_{\mathcal{R}} k_{y \to y'} x^y (y'-y),
		\end{equation}
for  $x \in \overline{\mathbb{R}}_+^{\mathcal{S}}$,
and its entries are called the \textbf{system polynomials} of $\mathcal{N}$.	
The \textbf{steady-state ideal} is $I_\mathcal{N} \coloneqq \langle f_1(x), \dotsc, f_n(x) \rangle \subseteq \mathbb{Q}[ \kappa; x]$, where $\mathbb{Q}[ \kappa; x]$ denotes the polynomial ring with variables $\{x_i\}_{i=1}^n$ and $\{\kappa_{y \to y'}\}_{\mathcal{R}}$.
The $\textbf{stoichiometric matrix}$ associated with $\mathcal{N}$ and denoted by $N$ is the $|\mathcal{S}| \times |\mathcal{R}|$ matrix with entries in $\mathbb{Z}$ where each column corresponds to the reaction vector $y \to y'$. 
We  call $\rank(N)$ the \textbf{rank of the network}.

A \textbf{conservation matrix} $W \in \mathbb{R}^{(n-s) \times n}$ is any rank $n-s$ matrix that satisfies $WN =0$, that is $W$ is a left kernel of $N$. The \textbf{corank} of $N$ is the rank of $W$. For each $c \in \mathbb{R}^{n-s}$ we have an associated \textbf{stoichiometric compatibility class} given by 
$\mathcal{P}_c \coloneqq \{x \in \mathbb{R}^n_{\geq 0} : Wx = c  \}$.
We note that $\mathcal{P}_c = \emptyset  $ if $c \notin W(\mathbb{R}^n_{\geq 0})$. The \textbf{positive stoichiometric compatibility class} is simply 
$\mathcal{P}_c^+ \coloneqq \mathcal{P}_c \cap \mathbb{R}^n_{>0}.$
	
		When considering trajectories, we will want to think of the paths taken by single points in the space $\mathbb{R}_+^{\mathcal{S}}$.  We refer to such points as a \textbf{composition}, denoted by $c$ or $c'$.
A composition $c \in \mathbb{R}^n$ can reach a composition $c' \in \mathbb{R}^n$ only if $c -c'$ lies in the same stoichiometric compatibility class.
For a mass-action kinetic system, a \textbf{steady state}, \textbf{fixed point} or \textbf{equilibrium} of the system is a point $c \in \mathbb{R}^{\mathcal{S}}$ such that $f(c) = 0.$  The set of \textbf{equilibria} is $V = \{c \in \mathbb{R}^n_{\geq 0} | f(c) =0 \}.$
A positive equilibrium $c^*$ is an equilibrium with the added restriction that $c^* \in \mathbb{R}_+^{\mathcal{S}}$. The set of \textbf{positive equilibria} is $V_+ =  \{c^* \in \mathbb{R}^n_+ | f(c^*) =0 \}.$
Retaining the notations above, for $\xi \in \mathbb{R}$, $x \in \mathbb{R}^n_{\geq 0}$, and $\omega$ a row of $W$, any equation of the form $\omega \cdot x = \xi$ is a \textbf{conservation relation}.  Then $\mathcal{N}$ is \textbf{conservative} if 
		$S \cap \mathbb{R}^{\mathcal{S}}_{+} \neq \emptyset.$

For algebraic definitions and constructions related to multivariate polynomials and Gro\"obner bases, we refer the reader to \cite{CLOfifth} for detailed background. We apply the theory of Gro\"obner bases in the polynomial ring $\mathbb{Q}[\kappa;x]$ described above, and any lexicographic order where the variable $x_1$ is greatest so that we can eliminate it.
	
	A vital question is if a reaction network has the capacity to admit exactly one steady state (monotstationary) or more than one steady state (multistationary), based solely on its structure.  Numerical approaches, including parameter estimation, are computationally costly in most cases of moderately sized or complicated reaction networks.  A theoretical approach that answers this more qualitative question would be very useful. Methods exist that can answer this question, without the need for numerical simulation or experimental data, but are inconclusive if certain conditions on the reaction network are not met. This is more frequent when the pathway for a phenomenon is not well understood, so that researchers are unsure about the fidelity of a proposed reaction network model for the phenomenon. Another issue with some of the methods is the reliance on the network to admit only specific kinetics for the model. Since we are focused on mass-action kinetics, we will mention when any of the conclusions are applicable to kinetics other than mass-action.
	
	Discussed at length in \cite{Feinberg_2019_book}, the deficiency-based theory is very strong. It has criteria that are easily checked, after which we can draw a conclusion depending on if the reaction network is weakly reversible or not. To discuss the theorem, we first need to introduce a few more definitions.
	
	Following \cite{Feinberg_2019_book}, we write $y \leftrightarrow y'$ if either $y \to y'$ or $y' \to y$.  We define an equivalence relation on $\mathcal{C}$ by setting $y\sim y'$ if $y = y'$ or there exists a sequence of complexes $y_1,\dotsc,y_k$ such that
			$$y = y_1 \leftrightarrow y_2 \leftrightarrow \cdots \leftrightarrow{}y_k = y'.$$
This partitions $\mathcal{C}$ into a set of \textbf{linkage classes} and we denote the number of linkage classes by $\ell$.
		
When looking at the dynamics of a reaction network, we care about the directions in which the reactions take place. 
Firstly, we say $y \rightarrow y'$ is \textbf{reversible} if also $y' \rightarrow y$.
We say that $y \in \mathcal{C}$ \textbf{ultimately reacts} to $y' \in \mathcal{C}$, written as $y \Rightarrow y'$, if either $y = y'$ or 
there exists a sequence of complexes $y_1,\dotsc,y_k$ such that
			$$y = y_1 \to y_2 \to \cdots \xrightarrow{}y_k = y'.$$
The equivalence relation on $\mathcal{C}$ defined by $y\sim y'$ when both $y \Rightarrow y' $ and $y' \Rightarrow y$ partitions $\mathcal{C}$ into \textbf{strong linkage classes}.
A reaction network $\mathcal{N}$ is \textbf{weakly reversible} if every linkage class of a reaction network is a strong linkage class.
We call a strong linkage class $\tau \subseteq \mathcal{C}$ a \textbf{terminal strong linkage class} if, for every complex $y \in \tau$ and any reaction $y \to y'$, we have $y' \in \tau$. We denote the number of terminal strong linkage classes by $t$.
	
		The \textbf{deficiency} of a reaction network with $|\mathcal{C}| =m$, $\rank(\mathcal{N})=s$, and $\ell$ linkage classes, is $\delta \coloneqq m -s -\ell$. The deficiency of a reaction network is a measure of how connected the network is: the lower the deficiency of a network, the fewer linear dependencies the network has. Deficiency-zero reaction networks are as linearly independent as the partition of linkage classes allows. Furthermore, any subnetwork of a deficiency-zero network will also have deficiency zero.
		
	One of the oldest and most important results in Deficiency Theory, which first appeared in the manuscripts by Feinberg \cite{deficiencyzero1}, Horn \cite{deficiencyzero2}, and Horn \& Jackson \cite{Horn_Jackson_1972}, is the Deficiency Zero theorem,
	
	\begin{theorem}[Feinberg, Horn, Jackson]
		\label{thm:deficiency zero}
		Let $\mathcal{N} = \{\mathcal{S}, \mathcal{C}, \mathcal{R}  \}$ be a reaction network such that $\delta = 0$. 
		\begin{enumerate}
			\item \label{thm:deficiency zero: first} If $\mathcal{N}$ is not weakly reversible, then for any kinetics $\mathcal{K}$, the set of positive equilibria is empty.
			\item \label{the:deficiency zero: second} If $\mathcal{N}$ is weakly reversible and has mass-action kinetics, then each positive stoichiometric class admits exactly one positive equilibrium. 
			Furthermore, this equilibrium is asymptotically stable.
		\end{enumerate}
	\end{theorem}

    If $s_i$ is the rank of linkage class $i$, then it can be shown that $s \leq \sum_{i=1}^{\ell} s_i$.   The deficiency of linkage class $i$ can then be written as $\delta_i = m_i -1 - s_i$, where $m_i$ is the number of complexes in linkage class $i$. 
    It can then be shown that
	
	\begin{equation}\label{eq:deficiency with linkage classes}
		\sum_{i=1}^{\ell} \delta_i = \sum_{i=1}^{\ell}(n_i - 1 - s_i) = n- \ell - \sum_{i=1}^{\ell}s_i \leq \delta.
	\end{equation}

    We get $\delta = \sum_{i=1}^{\ell} \delta_i$ exactly when $s = \sum_{i=1}^{\ell} s_i$,
	which can be thought of as the linkage classes being as independent from one another as possible. For example, this occurs when gluing two disjoint reaction networks, as described in the next section.
	

	\section{Theoretical results on joining and decomposing chemical reaction networks}
	
	In systems biology we often find distinct models trying to describe the same phenomenon. This could happen, for example, when incomplete data on the phenomenon is collected. When using mass-action kinetics, one way to distinguish between models is by comparing the Gr\"{o}bner bases of the ideals that are generated by the system polynomials associated with each model. The biggest issue with this approach is that computational complexity grows very quickly as the number of species increases, and computing Gr\"{o}bner bases can be infeasible even for moderately sized networks (between 10 and 20 species). Increased connectivity also increases the computational cost.
	
	We can decrease computational cost by projecting down to smaller networks within the original network of interest.
This is related to decomposing a reaction network into subnetworks, and can be approached in several different ways. We focus on gluing and cutting reaction networks as discussed in \cite{Gross_Harrington_Meshkat_Shiu_2020}. 
Motivation for this sort of approach comes from biology; for example in the phenomenon of cross-talk, which is when two (or more) distinct pathways interact with one another via the presence of a shared molecule. Thus it is important to develop a framework in reaction network theory that allows for the joining and decomposing of reaction networks.
	
	Consider two reaction networks $\mathcal{N}_1 = \{\mathcal{S}_1, \mathcal{C}_1,\mathcal{R}_1 \}$ and $\mathcal{N}_2 = \{\mathcal{S}_2, \mathcal{C}_2,\mathcal{R}_2 \}$. Their \textbf{union} is

	\begin{equation}
		\mathcal{N}_1 \cup \mathcal{N}_2 = \{\mathcal{S}_1 \cup \mathcal{S}_2, \mathcal{C}_1 \cup \mathcal{C}_2, \mathcal{R}_1 \cup \mathcal{R}_2\}.
	\end{equation}
	If $\mathcal{S}_1 \cap \mathcal{S}_2 = \emptyset$ for $\mathcal{N}_1, \mathcal{N}_2$, then the networks are completely disjoint and we can completely analyze the  
	union of the two networks by analyzing each subnetwork separately. There are several more complicated scenarios, as described in \cite{Gross_Harrington_Meshkat_Shiu_2020}; we focus on the following in this article.

\begin{definition} In the setup above, suppose that the two networks have at least one non-zero complex in common ($\mathcal{C}_1 \cap \mathcal{C}_2 \nsubseteq \{0\}$), but no reactions in common ($\mathcal{R}_1 \cap \mathcal{R}_2 = \emptyset$).  Then we say that the union $\mathcal{N}_1 \cup \mathcal{N}_2$ is obtained by \textbf{gluing over complexes}.
\end{definition}

An issue that can occur in analyzing reaction networks is that some species concentrations of a system may not be experimentally measurable.
Algebraically, this is approached by eliminating variables to restrict our attention to the subset of species variables which are measurable.  Given a subset of species for a reaction network  $\{j_1,j_2,\dotsc, j_l\}  \subset \mathcal{S}$, whose concentrations are measurable, we define the variables $x^{\text{obs}} := \{x_{j_1},x_{j_2},..., x_{j_l} \}$ and 
	\begin{equation}
		I_{\mathcal{N}}^{\text{obs}} := I_{\mathcal{N}} \cap \mathbb{Q}[\kappa;  x^{\text{obs}} ].
	\end{equation}
This is known as an \textbf{elimination ideal} in Gr\"{o}bner basis theory, and describes the steady states of the reaction network only in terms of species whose concentrations are measurable (or observable).

	Our main approach is to understand the algebraic interaction between the operations of projecting to subnetworks in order to reduce computational cost, and eliminate nonmeasurable species. To be precise, we define the following projection maps.
	
Let $\kappa(i)$ be the vector of indeterminate rate constants for the network $\mathcal{N}_i$. Similarly, let $x(i)$ be the subset of variables indexed by $\mathcal{S}_i$, and let $I_{\mathcal{N}_i}^{\text{obs}} = I_{\mathcal{N}_i} \cap \mathbb{Q}[\kappa(i);  x^{\text{obs}} ],$ where $i = 1 , 2$. To go from $I_{\mathcal{N}}$ to $I_{\mathcal{N}_i}$ we have the following projection ring homomorphism which corresponds to setting the concentrations of species outside of $\mathcal{N}_i$, as well as the rate constants corresponding to reactions outside of $\mathcal{N}_i$, all equal to zero.	
	\begin{subequations}
	\begin{align}\label{eqn:elimination}
		\phi_i : \mathbb{Q}[\kappa;x] & \to \mathbb{Q}[\kappa(i);x(i)];  \\
		\kappa_a & \mapsto 
		\begin{cases}
			\kappa_a, & \text{if } \kappa_a \in \kappa(i),\\
			0, & \text{if } \kappa_a \notin \kappa(i);
		\end{cases} 
		\\
		x_a & \mapsto 
		\begin{cases}
			x_a, & \text{if } x_a \in x(i);\\
			0, & \text{if } x_a \notin x(i).
		\end{cases}
	\end{align}
	\end{subequations}
This ring homomorphism is surjective by construction, and for a polynomial $f \in \mathbb{Q}[\kappa(i); x(i)]$, we have $f \in \mathbb{Q}[\kappa ; x]$ and $\phi_i(f) = f$.
	
Experimentally,	the subset of species whose concentrations are not observable may be distributed in any way across the smaller networks  $\mathcal{N}_1$ and  $\mathcal{N}_2$. The best case scenario is when projection onto each of the subnetworks commutes with the elimination of nonobservable species, which algebraically can be expressed by
		\begin{equation}\label{eq:mainequality}
			\phi_i (I_{\mathcal{N}}^{\text{obs}}) = I_{\mathcal{N}_i}^{\text{obs}},  \qquad \text{for } i=1,2.
		\end{equation}
	
	Gross, Harrington, Meshkat, and Shiu \cite{Gross_Harrington_Meshkat_Shiu_2020} showed that if we glue $\mathcal{N}_1$ and $\mathcal{N}_2$ over a set of complexes or a set of reactions, then we always have
		\begin{equation} \label{eq:equivalent ideals before elimination}
			I_{\mathcal{N}_i} = \phi_i(I_{\mathcal{N}}) \subseteq \mathbb{Q}[\kappa(i);x(i)], \qquad i=1,2,
		\end{equation}
	and furthermore that
	\begin{equation} \label{eq:elimcommute containment}
		\phi_i (I_{\mathcal{N}} \cap \mathbb{Q}[\kappa ; x^{\text{obs}}]) \subseteq I_{\mathcal{N}_i} \cap \mathbb{Q}[\kappa(i); x^{\text{obs}}], \qquad i=1,2.
	\end{equation}
In other words, in Equation \eqref{eq:mainequality} we always have the containment $\phi_i (I_{\mathcal{N}}^{\text{obs}}) \subseteq I_{\mathcal{N}_i}^{\text{obs}}$.
Our main result of this section is that the reverse containment holds for a specific family of reaction networks. We begin with an example that shows that equality does not always hold in Equation \eqref{eq:mainequality}.

In the following example and for the remainder of the paper, when we eliminate a single species $X_i$ from a reaction network so that $x^{\text{obs}} = \{x_1,...,x_{i-1},x_{i+1},...x_n \}$, we adopt the shorthand 
$$ I_{\mathcal{N}}^{\text{obs}} := I_{\mathcal{N}} \cap \mathbb{Q}[\kappa;  x^{\text{obs}} ] = I_{\mathcal{N}}^{x_i}.$$
That is, $I_{\mathcal{N}}^{x_i}$ is the steady-state ideal of the full network $\mathcal{N}$ once we eliminate species $X_i$.
	
	\begin{eg}\label{eg:2 species cyclic RN graph}  Here we present a situation in which equality of ideals does not hold.
		
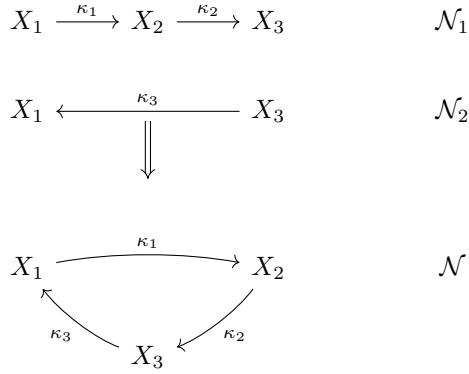
\begin{center}
\[\begin{tikzcd}
	{X_1} & {X_2} & {X_3} && {\mathcal{N}_1} \\
	{X_1} && {X_3} && {\mathcal{N}_2} \\
	& {} \\
	{X_1} && {X_2} && {\mathcal{N}} \\
	& {X_3}
	\arrow["{\kappa_1}", from=1-1, to=1-2]
	\arrow["{\kappa_2}", from=1-2, to=1-3]
	\arrow[""{name=0, anchor=center, inner sep=0}, "{\kappa_3}"', from=2-3, to=2-1]
	\arrow["{\kappa_1}", curve={height=-6pt}, from=4-1, to=4-3]
	\arrow["{\kappa_2}", curve={height=-6pt}, from=4-3, to=5-2]
	\arrow["{\kappa_3}", curve={height=-6pt}, from=5-2, to=4-1]
	\arrow[shorten <=3pt, Rightarrow, from=0, to=3-2]
\end{tikzcd}\]		\end{center}	
	
	Let $\mathcal{N}_1$ be given by $X_1 \to X_2 \to X_3$, $\mathcal{N}_2$ be given by $X_3 \to X_1$, and $\mathcal{N} = \mathcal{N}_1 \cup \mathcal{N}_2$ be obtained by gluing over complexes $X_1$ and $X_3$. We have the following system of ODEs for $\mathcal{N}$,
	\begin{subequations}
	\begin{align} \label{eg:2 species cyclic RN ODE N}
		\dot{x}_1 &= -\kappa_1x_1 + \kappa_3x_3,  \\
		\dot{x}_2 &= \kappa_1x_1 - \kappa_2x_2 ,  \\
                \dot{x}_3 &= \kappa_2x_2 - \kappa_3x_3.
	\end{align}
	\end{subequations}
	
	For the subnetwork $\mathcal{N}_1$, we have
	\begin{subequations}
	\begin{align} \label{eg:2 species cyclic RN ODE N1}
		\dot{x}_1 &= -\kappa_1x_1 ,\\
		\dot{x}_2 &= \kappa_1x_1 -\kappa_2x_2, \\
                \dot{x}_3 &= \kappa_2x_2.
	\end{align}
	\end{subequations}
	
	For the subnetwork $\mathcal{N}_2$, we have
	\begin{subequations}
	\begin{align} \label{eg:2 species cyclic RN ODE N2}
		\dot{x}_1 &= \kappa_3x_3, \\
		\dot{x}_2 &= 0, \\
                \dot{x}_3 &= -\kappa_3x_3 = -\dot{x}_1
	\end{align}
	\end{subequations}
	
	Thus we have the following steady-state ideals:
	\begin{subequations}
	\begin{align} \label{eg:2 species cyclic RN steady state ideals}
		I_{\mathcal{N}} &= \langle -\kappa_1x_1 + \kappa_2x_2, -\kappa_2x_2 + \kappa_3x_3 \rangle, \\
		I_{\mathcal{N}_1} &= \langle \kappa_1x_1, \kappa_2x_2  \rangle,  \\
		I_{\mathcal{N}_2} &= \langle  \kappa_3x_3 \rangle.
	\end{align}
	\end{subequations}
	
	If we eliminate species $X_3$, we get the following elimination ideals:
	\begin{subequations}
	\begin{align} \label{eg:2 species cyclic RN eliminated steady state ideals}
		I_{\mathcal{N}}^{x_3} &= \langle -\kappa_1x_1 + \kappa_2x_2 \rangle ,\\
		I_{\mathcal{N}_1}^{x_3} &= \langle \kappa_1x_1, \kappa_2x_2 \rangle , \\
		I_{\mathcal{N}_2}^{x_3} &= \langle  0 \rangle.
	\end{align}
	\end{subequations}
	
	Thus we have inequality of the ideals,
	\begin{equation}
		\phi_1 (I_{\mathcal{N}}^{x_3}) = \langle -\kappa_1x_1 + \kappa_2x_2   \rangle  \neq \langle \kappa_1x_1, \kappa_2x_2 \rangle = I_{\mathcal{N}_1}^{x_3},
	\end{equation} 
	showing that elimination of species does not always commute with projection to a subnetwork. Note that if we eliminated $x_1$ or $x_2$, the projection and elimination would commute. 
	\end{eg}	
	
We now introduce the terminology needed to state the main result of this section.
	
\begin{definition}\label{def:simple RN}
	A \textbf{simple} reaction network is one where $y \to y'$ implies that $(y', y) \notin \mathcal{R}$, {\em i.e.,} none of the reactions in the network are reversible. 
\end{definition} 

\begin{definition}\label{def:cyclic RN}
	A \textbf{cyclic} reaction network $\mathcal{N} = \{\mathcal{S}, \mathcal{C}, \mathcal{R} \}$ is one where the graph of the reaction network is a chordless cyclic graph. 
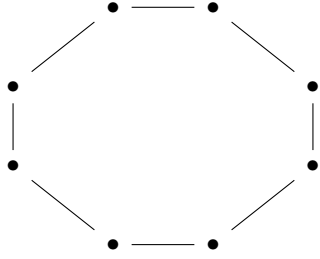
\begin{center}
\[\begin{tikzcd}[ampersand replacement=\&]
	\& \bullet \& \bullet \\
	\bullet \&\&\& \bullet \\
	\bullet \&\&\& \bullet \\
	\& \bullet \& \bullet
	\arrow[no head, from=1-2, to=1-3]
	\arrow[no head, from=1-3, to=2-4]
	\arrow[no head, from=2-1, to=1-2]
	\arrow[no head, from=2-4, to=3-4]
	\arrow[no head, from=3-1, to=2-1]
	\arrow[no head, from=3-4, to=4-3]
	\arrow[no head, from=4-2, to=3-1]
	\arrow[no head, from=4-3, to=4-2]
\end{tikzcd}\]
\end{center}
That is, for every $y \in \mathcal{C}$ we have $y \leftrightarrow y_1 \leftrightarrow ... \leftrightarrow y_i \leftrightarrow y$ where $y_i$ cycles through all of $\mathcal{C}$ and none of the $y_i$ repeat. 
	\end{definition} 
		
We can now prove the main result of this section.
	
	\begin{theorem} \label{result:theorem 2 complex cuts}
		Let $\mathcal{N} = \{\mathcal{S}, \mathcal{C}, \mathcal{R}\}$ be a  monomolecular, cyclic, simple, weakly-reversible mass-action kinetic system with $|\mathcal{S}| \geq 4$. If we cut $\mathcal{N}$ along any two complexes and form $\mathcal{N}_1 = \{\mathcal{S}_1, \mathcal{C}_1, \mathcal{R}_1 \}$ and $\mathcal{N}_2 = \{\mathcal{S}_2, \mathcal{C}_2, \mathcal{R}_2 \}$ such that $\mathcal{N} = \mathcal{N}_1 \cup \mathcal{N}_2$, where $|\mathcal{S}_1| \geq 3$ and $|\mathcal{S}_2| \geq 3$, then elimination of any species $X_j$ commutes with projection onto the subnetworks:
		\begin{equation}
			\phi_i (I_\mathcal{N} ^{x_j}) = I_{\mathcal{N}_i}^{x_j},   \qquad i = 1, 2.
		\end{equation}
	\end{theorem}
	
	\begin{proof}
		
We choose labels so that our network is of the following form and $X_1$ is the target for elimination,
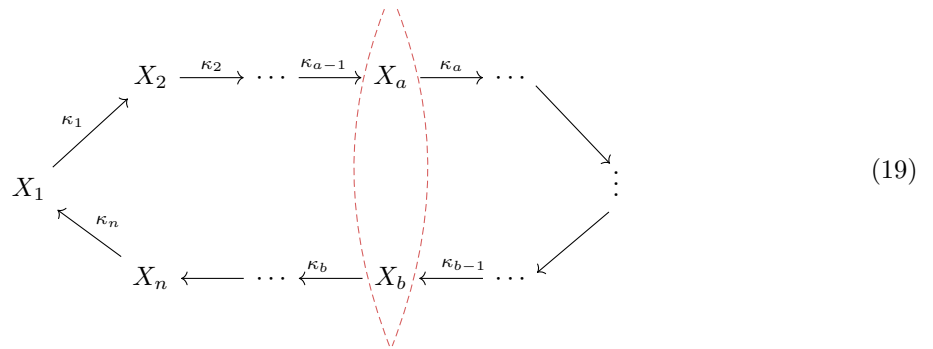
\begin{equation} \label{result:theorem example RN pre cut}
\begin{tikzcd}
	&&& {} \\
	& {X_2} & \cdots & {X_a} & \cdots \\
	{X_1} &&&&& \vdots \\
	& {X_n} & \cdots & {X_b} & \cdots \\
	&&& {}
	\arrow[color={rgb,255:red,214;green,92;blue,92}, curve={height=-18pt}, dashed, no head, from=1-4, to=5-4]
	\arrow["{\kappa_2}", from=2-2, to=2-3]
	\arrow["{\kappa_{a-1}}"{pos=0.4}, from=2-3, to=2-4]
	\arrow["{\kappa_a}", from=2-4, to=2-5]
	\arrow[shift left=2, from=2-5, to=3-6]
	\arrow["{\kappa_1}", from=3-1, to=2-2]
	\arrow[shift left=2, from=3-6, to=4-5]
	\arrow["{\kappa_n}"', from=4-2, to=3-1]
	\arrow[from=4-3, to=4-2]
	\arrow["{\kappa_b}"'{pos=0.7}, from=4-4, to=4-3]
	\arrow["{\kappa_{b-1}}"'{pos=0.3}, from=4-5, to=4-4]
	\arrow[color={rgb,255:red,214;green,92;blue,92}, curve={height=-18pt}, dashed, no head, from=5-4, to=1-4]
\end{tikzcd}
\end{equation}
with cuts along the complexes $X_a$ and $X_b$, where $1 \leq a< b-1\leq n-1$ (because of our assumptions that $|\mathcal{S}| \geq 4$, $|\mathcal{S}_1| \geq 3$, $|\mathcal{S}_2| \geq 3$). Therefore, $\mathcal{N}_1$ contains the complexes $X_a, X_{a+1}, \dotsc, X_b$, and $\mathcal{N}_2$ contains the complexes $X_b, X_{b+1}, \dotsc, X_n, X_1, \dotsc, X_a$.

We begin by finding a Gr\"obner basis of each ideal from which we need to eliminate $x_1$.  In this particular orientation, it turns out that the system polynomials for $\mathcal{N}$,
\begin{equation}
-\kappa_1 x_1 + \kappa_n x_n,\quad \kappa_1 x_1 - \kappa_2 x_2,\quad \kappa_2 x_2 - \kappa_3 x_3,\quad \dotsc,\quad \kappa_{n-1} x_{n-1} - \kappa_n x_n,
\end{equation}
already form a Gr\"obner basis. This is because the only terms appearing are of the form $\kappa_i x_i$, so the leading terms of the system polynomials are always equal or relatively prime, which guarantees that $S$-pairs between the system polynomials already reduce to zero in Buchberger's algorithm (see \cite[\S2.9~Prop.~4]{CLOfifth}).

We find Gr\"obner bases for $I_{\mathcal{N}_1}$ and $I_{\mathcal{N}_2}$ by noting that they are actually quadratic monomial ideals, so a set of monomial generators is already a Gr\"obner basis.  The system polynomials for $\mathcal{N}_1$ are
\begin{equation}
-\kappa_a x_a, \quad \kappa_a x_a - \kappa_{a+1} x_{a+1}, \quad \dotsc,\quad \kappa_{b-2} x_{b-2} - \kappa_{b-1} x_{b-1}, \quad \kappa_{b-1} x_{b-1}.
\end{equation}
Collapsing telescoping sums from these generators gives
\begin{equation}
I_{\mathcal{N}_1} = \langle \kappa_a x_a, \dotsc, \kappa_{b-1} x_{b-1}\rangle
\end{equation}
The system polynomials for $\mathcal{N}_2$ are, for $a=1$,
\begin{subequations}
\begin{equation}
-\kappa_b x_b, \ \kappa_b x_b - \kappa_{b+1} x_{b+1}, \dotsc, \ \kappa_{n-1} x_{n-1} - \kappa_{n} x_{n}, \ \kappa_{n} x_{n}.
\end{equation}
For $a=2$ the system polynomials for $\mathcal{N}_2$ are 
\begin{equation}
\kappa_1x_1, \dotsc,\ \kappa_{b} x_{b} - \kappa_{b-1} x_{b-1}, \ \kappa_{b-1} x_{b-1}.
\end{equation}
For $a=3$ the system polynomials for $\mathcal{N}_2$ are 
\begin{equation}
\kappa_1x_1-\kappa_2 x_2, \dotsc, \ \kappa_{a-2} x_{a-2} - \kappa_{a-1} x_{a-1},\ \kappa_{a-1} x_{a-1}, \kappa_b x_b,\  \kappa_b x_b - \kappa_{b+1}x_{b+1},\dotsc,\ \kappa_{n-1} x_{n-1} - \kappa_{n} x_{n}.
\end{equation}
\end{subequations}
Collapsing telescoping sums from these generators gives
\begin{equation}
I_{\mathcal{N}_2} =
\begin{cases}
\langle \kappa_{b} x_{b}, \dotsc, \kappa_n x_n \rangle, & \text{if }  a=1,\\
\langle \kappa_1 x_1, \dotsc, \kappa_{a-1} x_{a-1},  \kappa_{b} x_{b}, \dotsc, \kappa_n x_n \rangle, & \text{if }  a\geq 2.\\
\end{cases}
\end{equation}
In each case, the monomial generators are indexed by reactions appearing on the right and left sides of the cuts in Diagram \eqref{result:theorem example RN pre cut}, respectively.

With Gr\"obner bases of all ideals of interest at our disposal, and that fact that lex order is an elimination order for $x_1$, we can eliminate $x_1$ from each by simply taking the ideal generated by the elements of the Gr\"obner where $x_1$ does not appear \cite[\S3.1~Thm.~2]{CLOfifth}.  We find
\begin{subequations}
\begin{align}
I_{\mathcal{N}}^{x_1} &=  \langle \kappa_2 x_2 - \kappa_3 x_3,\ \dotsc,\ \kappa_{n-1} x_{n-1} - \kappa_n x_n \rangle \\
I_{\mathcal{N}_1}^{x_1} &= 
\begin{cases}
\langle \kappa_{2} x_{2} , \dotsc, \kappa_{b-1} x_{b-1} \rangle & \text{if }  a=1\\
\langle \kappa_a x_a, \dotsc, \kappa_{b-1} x_{b-1} \rangle & \text{if }  a\geq 2\\
\end{cases}\label{eq:IN1}\\
I_{\mathcal{N}_2}^{x_1} &= 
\begin{cases}
\langle  \kappa_{b} x_{b}, \dotsc, \kappa_n x_n \rangle & \text{if }  a=1, 2\\
\langle \kappa_2 x_2, \dotsc, \kappa_{a-1} x_{a-1},  \kappa_{b} x_{b}, \dotsc, \kappa_n x_n \rangle & \text{if }  a\geq 3\\
\end{cases}\label{eq:IN2}
\end{align}
\end{subequations}

We still need to compute the projections $\phi_i(I_{\mathcal{N}}^{x_1})$ for $i=1,2$. For $i=1$, we find that 
\begin{equation}
\phi_1(I_{\mathcal{N}}^{x_1}) = 
\begin{cases}
\langle \kappa_2 x_2 - \kappa_3 x_3,\ \dotsc,\ \kappa_{b-1} x_{b-1} - 0 \rangle,  & \text{if }  a=1, 2, \\
\langle 0 - \kappa_a x_a,\ \kappa_a x_a - \kappa_{a+1} x_{a+1},\ \dotsc,\ \kappa_{b-1} x_{b-1} - 0 \rangle  & \text{if }  a\geq 3,
\end{cases}
\end{equation}
and by collapsing telescoping sums of generators we see that this is equal to Equation \eqref{eq:IN1} in all cases.
For $i=2$, we find that
\begin{equation}
\phi_2(I_{\mathcal{N}}^{x_1}) = 
\begin{cases}
\langle  0 - \kappa_bx_b,\ \dotsc,\ \kappa_{n-1} x_{n-1} - \kappa_n x_n\rangle, & \text{if }  a=1,2,\\
\langle \kappa_2 x_2 - 0, 0 - \kappa_bx_b,\ \dotsc,\ \kappa_{n-1} x_{n-1} - \kappa_n x_n\rangle, & \text{if }  a=3,\\
\langle \kappa_2 x_2 - \kappa_3 x_3,\ \dotsc,\ \kappa_{a-1} x_{a-1} - 0, 0 - \kappa_bx_b,\ \dotsc,\ \kappa_{n-1} x_{n-1} - \kappa_n x_n\rangle, & \text{if }  a\geq 4.
\end{cases}
\end{equation}
and again by collapsing telescoping sums of generators this is equal to Equation \eqref{eq:IN2} in all cases, completing the proof.
	\end{proof}

\begin{remark}\label{remark:conjecture}
	Future directions based on the results in this section include exploring whether the result generalizes to monomolecular, cyclic, simple reaction networks which are not weakly reversible.  An immediate challenge that one encounters in this case is that the system polynomials are not necessarily a Gr\"obner basis, and even in small examples a Gr\"obner basis may no longer be quadratic.
\end{remark}
	
	\section{Computations on multistationarity for biochemical reaction networks}
	
	In this section we examine a reaction network's capacity to admit multiple steady states. For what assignment of rate parameters (if ever) does a particular reaction network exhibit multiple equilibria?  This is addressed in many different ways. Classical results from Deficiency Theory include the  deficiency-zero and deficiency-one theorems in \cite{deficiencyzero1, deficiencyone, deficiencyzero2, Horn_Jackson_1972}. 
	There are also results based on sign criterion for the determinant of the Jacobian of the species-formation rate function  \cite{homotopymultistationarity2008, simplified_jc2012}. This is a bit more sophisticated, but it only tells us when a network is not multistationary. 
	
	We employ a method that involves analyzing the sign of each term of a polynomial that is the determinant of a special matrix, introduced in \cite{multistationarity_regions2017}. The main idea revolves around a theorem whose proof relies on Brower degree theory. 
	
	We consider two reaction networks: a gut bacteria network adapted from \cite{Adrian_Ayati_Mangalam_2023}, and an altered model of post-translational modification (PTM) of proteins.  We will use the algorithm from \cite{multistationarity_regions2017}, as well as Deficiency Theory, to find out when (or if) these modified models exhibit multistationarity for some assignment of parameter values. 
	
	We introduce the necessary theory and algorithm used for our main results on the PTM-of-proteins model. 
	
	\subsection{Algorithm from Conradi, Feliu, Mincheva and Wiuf 2017} \label{sec:algorithm} 
	
	Conradi, Feliu, Mincheva and Wiuf \cite{multistationarity_regions2017} showed that, for a dissipative reaction network $\mathcal{N}$ with stoichiometric matrix $N$, $s = \rank(N)$, $d = \rank(\mathcal{N})$, stoichiometric compatibility class $\mathcal{P}_c$ with no boundary equilibria and such that $\mathcal{P}_c^+ \neq \emptyset$ with $c \in \mathbb{R}^d$, and appropriate kinetics\footnote{Their result is shown for more general kinetics than mass action, though we will only look at mass action kinetics.}, the following holds:
		
		\begin{enumerate}
			\item If 
			\begin{equation} \label{thm:feliu multistationarity 2017:monostationary condition}
				\text{sign}(\det(M(x))) = (-1)^s 
			\end{equation}
			for all positive equilibria $x \in V \cap \mathcal{P}_c^+$, then there is exactly one positive equilibrium in $\mathcal{P}_c$ and this equilibrium is nondegenerate. 
			\item If 
			\begin{equation} \label{thm:feliu multistationarity 2017:multistationarity condition}
				\text{sign}(\det(M(x))) = (-1)^{s+1} 
			\end{equation}
			for some positive equilibrium $x \in V \cap \mathcal{P}_c^+$, then there are at least two distinct positive equilibria in $\mathcal{P}_c^+$ where at least one is nondegenerate. If all positive equilibria of $\mathcal{P}_c$ are nondegenerate then there are at least three distinct positive equilibria and the total number of positive equilibria will be odd.
		\end{enumerate}
	
	Note that Equation \eqref{thm:feliu multistationarity 2017:monostationary condition} is a condition for all positive equilibria, while Equation \eqref{thm:feliu multistationarity 2017:multistationarity condition} is just for some positive equilibrium.
	
	To help understand the algorithm we apply it to a modified Michaelis-Menten mechanism for the 1-site phosphorylation and dephosphorylation cycle, the so-called PTM-for-proteins model. Previous work was done on the n-site phosphorylation and dephosphorylation cycle reaction network but the location of a reversible reaction is switched in the network analysis we present here\footnote{Skye Dore-Hall, Jos\'{e} Lozano, and JMS worked on this example as part of the 2023 MSRI-MPI Leipzig Summer Graduate School on Algebraic Methods for Biochemical Reaction Networks.}. The widely studied $1$-site phosphorylation dephosphorylation cycle reaction network has the graph	
	
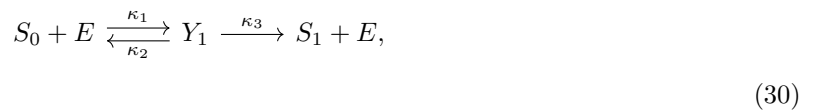
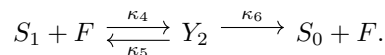
\begin{equation}
		\begin{tikzcd} \label{RN:1 site phosphorylation dephosphorylation standard}
			{S_0 + E} & {Y_1} & {S_1 +E,} \\
			\\
			{S_1+F} & {Y_2} & {S_0+F}
			\arrow["{\kappa_{1}}", shift left, from=1-1, to=1-2]
			\arrow["{\kappa_{3}}", from=1-2, to=1-3]
			\arrow["{\kappa_4}", shift left, from=3-1, to=3-2]
			\arrow["{\kappa_6}", shift left, from=3-2, to=3-3]
			\arrow["{\kappa_5}", shift left, from=3-2, to=3-1]
			\arrow["{\kappa_2}", shift left, from=1-2, to=1-1].
		\end{tikzcd}
	\end{equation}
	
	The network analysis presented here has the graph	
	
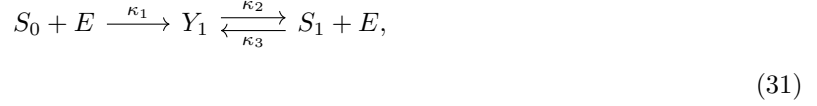
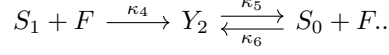
\begin{equation}
		\begin{tikzcd} \label{RN:1 site phosphorylation dephosphorylation modified}
			{S_0 + E} & {Y_1} & {S_1 +E,} \\
			\\
			{S_1+F} & {Y_2} & {S_0+F.}
			\arrow["{\kappa_{1}}", from=1-1, to=1-2]
			\arrow["{\kappa_{3}}", shift left, from=1-3, to=1-2]
			\arrow["{\kappa_4}", from=3-1, to=3-2]
			\arrow["{\kappa_5}", shift left, from=3-2, to=3-3]
			\arrow["{\kappa_2}", shift left, from=1-2, to=1-3]
			\arrow["{\kappa_6}", shift left, from=3-3, to=3-2].
		\end{tikzcd}
	\end{equation}
	
	The goal is to determine if the reaction network has the capacity for multistationarity. If the network does have the capacity to admit multiple steady states, we would like to know where in the parameter space this occurs. Computations were run for $n=1,2$. Although the framework is much the same for $n>2$, the symbolic computations involved do not scale well.

	\subsection{Steps of the algorithm} \label{subsection:1 site phospho dephospho}
	
	The algorithm can be divided into seven steps:
	
	\begin{enumerate} \label{def:algorithm}
		\item Find the system polynomials $f(x)$, the associated stoichiometric matrix $N$, and the row reduced echelon form of the conservation matrix $W$ for the network.
		\item Check if the network is dissipative.
		\item Make sure there are no relevant boundary steady states ({\em i.e.,} $X_i = 0$ for some $i$ at steady state) for the system.
		\item Construct $\phi_c(x)$ from $W$ and $f(x)$, then compute its Jacobian $M(x)$.
		\item Compute the polynomial $\det(M(x))$ and analyze the signs of each term. At this step, we can make a conclusion only if each term of $\det(M(x))$ has the same sign. Otherwise continue to step $6$.
		\item Find a positive parametrization $\psi$ of the set of positive steady states. Plug this into $M(x)$ and compute $\det(M(\psi(\Tilde{x})))$.
		\item Analyze the signs of the terms of $\det(M(\psi(\Tilde{x})))$, which will be a rational function. If the polynomial in the numerator has the same sign for all the terms, then conclude.  If it has 
		terms of different signs then compute its Newton polytope and check that the monomial with a sign satisfying Equation \eqref{thm:feliu multistationarity 2017:multistationarity condition} is a vertex for the Newton polytope. 
	\end{enumerate}
	
	We now discuss the first five steps in the context of the $1$-site phosphorylation dephosphorylation cycle reaction network.
	
	\begin{description}
		\item[Step 1:] Put $W$ in reduced row-echelon form. 
		
		For the network in Equation \eqref{RN:1 site phosphorylation dephosphorylation modified} we write out the system polynomials:		
		\begin{subequations}\label{eg: 1 site phosphorylation dephosphorylation modified ODE}
			\begin{align}
				\dot{s_0} &=  -k_1s_0e+k_5y_2-k_6s_0f, \\
				\dot{s_1} &=  k_2y_1-k_3s_1e-k_4s_1f, \\
				\dot{y_1}   &=  k_1s_0e-k_2y_1 + k_3s_1e, \\
				\dot{y_2}  &=  k_4s_1f -k_5y_2 + k_6s_0f, \\
				\dot{e}   &=  -k_1s_0e+k_2y_1-k_3s_1e, \\
				\dot{f}   &=  -k_4s_1f+k_5y_2 -k_6s_0f.
			\end{align}
		\end{subequations}
		
		Next we construct the stoichiometric matrix $N$, keeping the convention of the species ordering to be consistent with the order in which the ODEs in Equations $\eqref{eg: 1 site phosphorylation dephosphorylation modified ODE}$ were given,		
		\begin{equation}
			N = \begin{pmatrix}
				-1 & 0 & 0 & 0 & 1 & -1 \\
				0 & 1 & -1 & -1 & 0 & 0 \\
				1 & -1 & 1 & 0 & 0 & 0 \\
				0 & 0 & 0 & 1 & -1 & 1 \\
				-1 & 1 & -1 & 0 & 0 & 0 \\
				0 & 0 & 0 & -1 & 1 & -1
			\end{pmatrix}.
		\end{equation}
		
		We have $\rank (N) = 3$, and thus $W \in \mathbb{R}^{3 \times 6}$, which leads us to conclude that there are three conservation laws. We see, either from observation or with the aid of a software package, that they correspond to		
		\begin{equation}
			W = \begin{pmatrix} \label{eg:MSRI 1 site W}
				1 & 1 & 1 & 1 & 0 & 0 \\
				0 & 0 & 1 & 0 & 1 & 0 \\
				0 & 0 & 0 & 1 & 0 & 1
			\end{pmatrix}.
		\end{equation}
		
		\item[Step 2:] Check the dissipativity of the network.  A network $\mathcal{N}$ is \textbf{dissipative} if for all $\mathcal{P}_c$ there exists a convex set such that the trajectories of $\mathcal{P}_c$ eventually enter the convex set.  This is a weaker condition than if a network is conservative. Horn and Jackson \cite{Horn_Jackson_1972} showed that a positive stoichiometric class is bounded if and only if the reaction network it belongs to is conservative.
		
		We see that if a reaction network is conservative, the convex set for the trajectories of $\mathcal{P}_c$ is the stoichiometric compatibility class itself, since all trajectories starting within it stay inside it. We can check if a network is conservative very quickly using the software package CRNT \cite{CRNT_toolbox}. 
		
		Since the network in Equation \ref{RN:1 site phosphorylation dephosphorylation modified} has the conservation matrix given by Equation \ref{eg:MSRI 1 site W}, then by adding all three conservation laws ({\em i.e.,} rows of $W$), we get a positive vector that is orthogonal to $N$ (since $WN = 0$ by definition). Thus we conclude that our network is conservative and thus dissipative.

		\item[Step 3:]  A \textbf{boundary equilibrium point} is simply an equilibrium point in which at least one of the species is absent, {\em i.e.,} there exists $i \in \mathbb{Z}_+$ with $X_i \in \mathcal{S}$ such that $X_i =0$. This condition is generally not easy to check even for relatively small networks. One could try to analyze the system polynomials and check for a contradiction on the nonnegativity of the parameters and the rest of the species. This is time consuming and the approach is not always fruitful. Instead we can use the idea of siphons introduced in \cite{siphons_2010}. A nonempty set $Z \subseteq \mathcal{S}$ is a $\textbf{siphon}$ if, for every $X_i \in Z$ such that $X_i \in \supp{(y')}$ for $y \to y' \in \mathcal{R}$, then, for some $k$, there is an $X_k \in Z$ such that $X_k \in \supp{(y)}$.  This is saying that if the species $X_i$ is an element of our siphon $Z$ and is part of the product complex of a reaction, then there exists some species $X_k$ also belonging to our siphon such that $X_k$ is part of our reactant complex for the same reaction.
		
		Siphons were first introduced as semilocking sets in \cite{Anderson_2008,Anderson_Shiu_2010}. We call a siphon $\textbf{minimal}$ when it does not properly contain any other siphon. The following result will help us check for the existence of boundary equilibria in reaction networks. Shiu and Sturmfels \cite{siphons_2010} showed that if $\mathcal{Z} = \{Z_1,...,Z_a\}$ is a set of minimal siphons for $\mathcal{N}$, and for every minimal siphon $Z_i \in \mathcal{Z}$ there exists a subset of species $\{ X_{i_1}, ..., X_{i_k} \} \subseteq Z_i$ and a conservation relation $\lambda_1X_{i_1} + ... + \lambda_kX_{i_k} = c$ for some positive $\lambda_1,...,\lambda_k$, then $\mathcal{N}$ has no boundary equilibria in any of the stoichiometric compatibility classes $\mathcal{P}_c$ with  $\mathcal{P}_c^+ \neq \emptyset$.
		
		This proposition is for stoichiometric compatibility classes $\mathcal{P}_c$ with  $\mathcal{P}_c^+ \neq \emptyset$. The result of the proposition is exactly what we want (since each stoichiometric compatibility class is a compact set) and comes without much additional computational cost.  In summary, the only condition to check is that each minimal siphon of the network contains the support of a positive conservation relation.
		
		Finding siphons can be done algorithmically and is presented in \cite{siphons_2010}. For small networks we can find the set of all siphons $\mathcal{Z}$ fairly quickly. 
		
		For the network defined in Equation \eqref{RN:1 site phosphorylation dephosphorylation modified} we have the following siphons:
		\begin{subequations} \label{eq:1 site siphons}
			\begin{align}
				\{S_0,S_1 Y_1,Y_2 \}   &= Z_1,\\
				\{E,Y_1 \}  &= Z_2, \label{eq:1 site siphons:minimal 1} \\
				\{F,Y_2 \}  &= Z_3, \label{eq:1 site siphons:minimal 2}\\
				\{S_1, Y_1, E \} &= Z_4, \\
				\{S_0, Y_2, F \}  &= Z_5.
			\end{align}
		\end{subequations}
		Note that Equations \eqref{eq:1 site siphons:minimal 1} and \eqref{eq:1 site siphons:minimal 2} are minimal (each contains two species), and correspond to rows 2 and 3 of the conservation matrix given by Equation \ref{eg:MSRI 1 site W}, which are conservation laws for the network. Thus $\mathcal{N}$ has no boundary equilibria in stoichiometric compatibility classes $\mathcal{P}_c$ that satisfy $\mathcal{P}_c^+ \neq \emptyset$.

		\item[Step 4:] The system polynomials of a mass-action kinetic system can sometimes be redundant in the sense that the steady-state ideal can be generated by a proper subset of the system polynomials. This is particularly the case whenever you have nonzero corank or $\rank (N) < |\mathcal{S}|$. By way of construction of the conservation relations we can see that some system polynomials are linear combinations of others. However, we would like to encode information about the linear dependence of the system polynomials. As such, we will introduce a new construction that takes both the linearly independent system polynomials and all conservation relations and denote it by $\phi_c(X) : \mathbb{R}^n_{\geq 0} : \mathbb{R}^n$. Recall that the $W$ matrix is not unique. In order to get a unique construction of $\phi_c(x)$ we first need need to put $W$ in reduced row-echelon form. For this construction let us assume that $W \in \mathbb{R}^{d \times n}$ is the conservation matrix in row echelon form and let $\{k_1,...,k_d\}$ be the indices of the first nonzero entry of each row of $W$. Then		
		\begin{equation} \label{def:phi_c(x)}
			\phi_c(x)_i \coloneqq  
			\begin{cases} 
				f_i(x), & i \notin \{k_1,...,k_d \}, \\
				(WX - c)_i, & i \in  \{k_1,...,k_d \}.
			\end{cases}
		\end{equation}		
		We can see, since the equations in $\phi_c(x)$ contain all the conservation relations that define $\mathcal{P}_c$, that
		
		\begin{equation} \label{def:multiple equilibria phi function}
			V \cap \mathcal{P}_c = \{x \in \mathbb{R}^n_{\geq 0} | \phi_c(x) = 0 \}. 
		\end{equation}		
		If there are multiple solutions to Equation \eqref{def:multiple equilibria phi function}, then we say that the network admits multiple steady states for the particular stoichiometric compatibility class given by $c$. 
		
		To construct $M(x)$ we simply take the Jacobian of $\phi_c(x)$,		
		\begin{equation} \label{def:M(x)}
			M(x) \coloneqq J_{\phi}(x).
		\end{equation}		
		Even though $\mathcal{P}_c$ is defined by some constant $c \in \mathbb{R}$, the matrix $M(x)$ does not depend on $c$. Indeed the $i$-th row of $M(x)$ where $i \in \{k_1,...,k_d \}$ corresponds exactly to the conservation relation, that is the $i$-th row of our conservation matrix $W$.  

		Let $W$ and $\phi_c(x)$ be defined by Equation \eqref{def:phi_c(x)}, let $M(x)$ be defined by Equation \eqref{def:M(x)}, and let $x^* \in V \cap \mathcal{P}_c$. We say that $x^*$ is a \textbf{nondegenerate equilibrium point} if $M(x^*)$ is a nonsingular matrix, that is $\det(M(x^*)) \neq 0$. The function $\phi_c(x)$ and the matrix $M(x)$ are easily constructed given a conservation matrix $W$ for some reaction network. Step 1 and Step 4 are quick to implement. 
		
		For the $1$-site phosphorylation dephosphorylation cycle we note that our conservation matrix $W$ is already in reduced row-echelon form. We need to express the conservation laws in terms of the species of the network, as is written in Equation \eqref{def:phi_c(x)}. Let $c = (c_1,c_2,c_3) \in \mathbb{R}^3_+$ denote the constants for our conservation laws. Since $W$ has leading $1$s at $s_0$, $y_1$, and $y_2$, we replace those ODEs with the conservation laws to obtain		
		\begin{equation}
			\phi_c(x) = \begin{pmatrix}
				s_0 + s_1 + y_1 + y_2 - c_1 \\
				\dot{s}_1 \\
				y_1 + e -c_2 \\
				y_2 + f -c_3 \\
				\dot{e} \\
				\dot{f}
			\end{pmatrix} = \begin{pmatrix}
				s_0 + s_1 + y_1 + y_2 - c_1 \\
				k_2y_1 -k_3s_1e -k_4s_1f \\
				y_1 + e -c_2 \\
				y_2 + f -c_3 \\
				-k_1s_0e +k_2y_2 -k_3s_1e \\
				-k_4s_1f +k_5y_2 -k_6s_0f
			\end{pmatrix}.
		\end{equation}		
		Taking the Jacobian of $\phi_c(x)$ gives us the matrix $M(x)$:		
		\begin{equation}
			M(x) = \begin{pmatrix}
				1 & 1 & 1 & 1 & 0 & 0 \\
				0 & -k_3e -k_4f & k_2 & 0 & -k_3s_1 & -k_4s_1 \\
				0 & 0 & 1 & 0 & 1 & 0 \\
				0 & 0 & 0 & 1 & 0 & 1 \\
				-k_1e & -k_3e & k_2 & 0 & -k_1s_0 -k_3s_1 & 0 \\
				-k_6f & -k_4f & 0 & k_5 & 0 & -k_4s_1 -k_6s_0
			\end{pmatrix}.
		\end{equation}

		\item[Step 5:] Calculating $\det(M(x))$ is not an easy task. Since $M(x)$ will be of size $|\mathcal{S}| \times |\mathcal{S}|$, the symbolic computation of the determinant could be quite costly. Another issue in this step is when the terms of the resulting polynomial of $\det(M(x))$ do not all have the same sign. Without knowing information about the equilibria in $V \cap \mathcal{P_c}^+$, we cannot conclude anything about the values that the parameters using our existing tools, and must assume when the polynomial $\det(M(x))$ has monomials of distinct signs.
		
We use the Julia packages 
		\begin{itemize}
		\item \texttt{Symbolics.jl} \cite{gowda2021high},
		\item \texttt{LinearAlgebra.jl} \cite{basejulia2017},
		\item \texttt{ModelingToolkit.jl} \cite{ma2021modelingtoolkit}
		\end{itemize} 
to compute $\det(M(x))$ symbolically. The resulting polynomial from the 1-site phosphorylation dephosphorylation is
		\begin{multline} \label{eg:1 site phospho dephospho:det of M}
			\det(M(x)) =  -  k_1 k_2 k_5e -  k_2 k_4 k_5f -  k_1 k_3 k_5e^{2} -  k_2 k_4 k_6f^{2} \\
			-  k_1 k_2 k_4e f  -  k_1 k_4 k_5e f -  k_1 k_3 k_5e s_0 -  k_1 k_2 k_4e s_1 -  k_1 k_3 k_5e s_1 \\ 
			-  k_1 k_3 k_4s_1^{2} e -  k_1 k_2 k_6e s_0  -  k_1 k_4 k_5f s_0 -  k_1 k_4 k_6s_0^{2} f -  k_3 k_4 k_5f s_1 \\ 
			-  k_2 k_4 k_6f s_0 -  k_2 k_4 k_6f s_1 -  k_1 k_3 k_6e^{2} s_0 -  k_1 k_3 k_4e^{2} s_1  -  k_1 k_4 k_6f^{2} s_0 \\
			-  k_3 k_4 k_6f^{2} s_1 -  k_1 k_3 k_6s_0^{2} e -  k_3 k_4 k_6s_1^{2} f  -  k_1 k_3 k_6e f s_0 - k_1 k_4 k_6e f  s_0 \\
			-  k_1 k_3 k_4e f s_1 -  k_3 k_4 k_6e f s_1 -  k_1 k_3 k_4e s_0 s_1  - k_1 k_3 k_6e  s_0 s_1 - k_1 k_4 k_6f  s_0 s_1  \\
			- k_3 k_4 k_6f  s_0 s_1 
		\end{multline}
		Note that all the signs are negative. Since $s = \rank(N) = 3$ and $(-1)^s = (-1)^3 = -1$ which coincides with the sign of our determinant, we can use Clause $1$ of Theorem $\ref{thm:feliu multistationarity 2017:monostationary condition}$ to conclude that the $1$-site phosphorylation dephosphorylation cycle from is monostationary for all stoichiometric compatibility classes with non-empty interior.
	\end{description}

	Steps 6 and 7 are not specific to our example.
	
	\begin{description}
		\item[Step 6:] A positive parametrization of the set of positive equilibria is needed to apply Corollary 2 from \cite{multistationarity_regions2017}. Using Equations \eqref{thm:feliu multistationarity 2017:monostationary condition} and \eqref{thm:feliu multistationarity 2017:multistationarity condition} requires knowledge about the steady states inside $V \cap \mathcal{P}_c^+$. If we are lucky enough that all signs of the polynomial given by $\det (M(x))$ are the same, then, regardless of the values of the parameters or the specific equilibrium (since both $k = (k_1,..,k_r)$ and $x \in V \cap \mathcal{P}_c^+$ are vectors with all positive coordinates), we can satisfy either Equation \eqref{thm:feliu multistationarity 2017:monostationary condition} or Equation \eqref{thm:feliu multistationarity 2017:multistationarity condition}. 
		
		Should any term of $\det (M(x))$ have an opposite sign, then we cannot exclude the possibility of the determinant changing signs for some $x^* \in V \cap \mathcal{P}_c^+$. Thus far $\det (M(x))$ has been a polynomial over all species concentrations of the system. We can restrict it to be a polynomial over the steady states of the system, and thus have a much smaller parameter space to analyze. 
		
		The set of positive equilibria $V \cap \mathbb{R}^n_+$ has a \textbf{positive parametrization} if there exists a surjective function,
			\begin{align} \label{def:positive parametrization}
				\psi : \mathbb{R}^m_+ & \to V \cap  \mathbb{R}^n_+, \\
				\tilde{x} = (\tilde{x}_1,...,\tilde{x}_m) & \mapsto (\psi_1(\tilde{x}), ... , \psi_n(\tilde{x})),
			\end{align}
			for some $m<n$ such that $\tilde{x}$ is the vector of free variables $(\tilde{x}_1,...,\tilde{x}_m)$. That is, we can express every variable $x_i$ in terms of just $\tilde{x}$ for $1\leq i \leq n$. We call a positive parametrization $\textbf{algebraic}$ if the components $\psi_i(\tilde{x})$ are rational functions.
	
			Usually the number of free variables $m$ corresponds to the corank of the network. We also note that we will get rational expressions when solving for an $x_i$ in an ODE at steady state since our kinetics are mass action.

		Since $x_i = \phi_i(\tilde{x})$ for all $i$, we can rewrite
		\begin{equation}
			V \cap \mathcal{P}_c^+ = \{\psi(\tilde{x})  | \Tilde{x}\in \mathbb{R}^m_+  \}.
		\end{equation}
		
		Thus we can see that the stoichiometric classes of the positive parametrization are given by		
		\begin{equation} 
			V \cap \mathcal{P}_c^+ = \{\psi(\tilde{x}) | \tilde{x} \in \mathbb{R}^m_+ \And c = W \psi(\tilde{x})   \}.
		\end{equation}
		What this reformulation allows us to do is consider Equations \eqref{thm:feliu multistationarity 2017:monostationary condition} and \eqref{thm:feliu multistationarity 2017:multistationarity condition} applied to $M(x)$, when evaluated at $\psi(\Tilde{x})$:
		\begin{equation}
			a(\Tilde{x}) = \det(M(\psi(\Tilde{x}))), \Tilde{x} \in \mathbb{R}^m_+.
		\end{equation}
		
		\item[Step 7:] We now find sign conditions on monomial terms of
		$\det(M(\psi(\Tilde{x})))$. In the unfortunate case when there remain sign differences in the terms of $\det(M(\psi(\Tilde{x})))$, we can still find conditions on $\psi(\Tilde{x})$ such that, at steady state, a particular sign will dominate the entire polynomial, $\det(M(\psi(\Tilde{x})))$. This is achieved by computing the Newton polytope for $\det(M(\psi(\Tilde{x})))$ and checking that the term with a sign difference is a vertex of the Newton polytope.
		
		For a multivariate polynomial $f(x) \in \mathbb{R}[x_1,...,x_n]$ such that 
			$$f(x) = \sum_{\alpha \in \mathbb{N}^n} c_{\alpha}x^{\alpha},$$ where only finitely many $c_{\alpha} \in \mathbb{R}$ are nonzero, the \textbf{Newton polytope} of $f$ is
		\begin{equation}\label{def:newton polytope}
			\mathscr{N}(f) \coloneqq \left\{\sum_{k}\beta_k\alpha_k : \sum_{k}\beta_k =1,  \ \beta_j \geq 0 \text{ for all } j \right\}. 
		\end{equation} 

		The Newton polytope is the convex hull of the exponents $\alpha$ of the monomials of $f(x)$. 
		
		From \cite{multistationarity_regions2017} we use the result that for
			$$f(x) = \sum_{\alpha \in \mathbb{N}^n}c_{\alpha}x^{\alpha}, $$
		and $\alpha'$ a vertex of $\mathcal{N}(f)$, there exists $x' \in \mathbb{R}_+^n$ such that 
			$$\text{sign}(f(x')) = \text{sign}(c_{\alpha '}).$$
		
		Thus if one of the monomial terms, say $\psi(\hat{x})$ in  $\det(M(\psi(\Tilde{x})))$, has different signs and it is a vertex of $\mathscr{N}(\det(M(\psi(\Tilde{x}))))$, then we know that for some $\psi(\Tilde{x})$,  $\text{sign}(\det(M(\psi(\Tilde{x})))) = \text{sign}(\psi(\hat{x}))$, which will indicate either monostationarity or multistationarity, depending on $\dim(N) = s$.	
\end{description}

We summarize the seven steps: 	
	\begin{enumerate} \label{section:algorithm_project3}
		\item Find $f(x)$ and the row-reduced conservation matrix $W$. Also find $v(x)$ and check that it vanishes whenever one of the reactant species is absent. 
		\item Check that $\mathcal{N}$ is a dissipative reaction network.
		\item Make sure there are no boundary equilibria in any of the stoichiometric compatibility classes with nonempty interiors.
		\item Construct  $\phi_c(x)$ and $M(x)$. 
		\item Compute $\det (M(x))$. At this step, make sure that there are conditions on the parameters such that $\det (M(x))$ satisfies a lemma in \cite{multistationarity_regions2017}. Otherwise continue to steps 6 and 7. 
		\item Find a positive parametrization of the set of positive steady states $\psi$. Compute $\det(M(\psi(\Tilde{x})))$.
		\item Find sign conditions on $\det(M(\psi(\tilde{x})))$, which ensure either monostationarity or multistationarity. Otherwise check that at least one of the sign changes occurs on the vertex of $\mathscr{N}(\det(M(\psi(\Tilde{x}))))$.
	\end{enumerate}
	
	\subsubsection{The 2-site phosphorylation dephosphorylation cycle reaction network} \label{section:2 site phospho dephospho}
	
	We now focus our attention to the case when $n=2$ in the 2-site phosphorylation dephosphorylation cycle reaction network. The graph of the network is
	
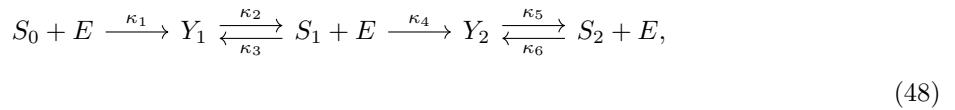
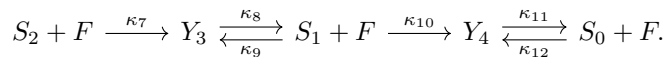
\begin{equation}
		\begin{tikzcd} \label{eg:2 site graph}
			{S_0 + E} & {Y_1} & {S_1 +E} & {Y_2} & {S_2+E,} \\
			\\
			{S_2+F} & {Y_3} & {S_1+F} & {Y_4} & {S_0+F.}
			\arrow["{\kappa_{1}}", from=1-1, to=1-2]
			\arrow["{\kappa_{3}}", shift left, from=1-3, to=1-2]
			\arrow["{\kappa_7}", from=3-1, to=3-2]
			\arrow["{\kappa_8}", shift left, from=3-2, to=3-3]
			\arrow["{\kappa_2}", shift left, from=1-2, to=1-3]
			\arrow["{\kappa_9}", shift left, from=3-3, to=3-2]
			\arrow["{\kappa_4}", from=1-3, to=1-4]
			\arrow["{\kappa_5}", shift left, from=1-4, to=1-5]
			\arrow["{\kappa_6}", shift left, from=1-5, to=1-4]
			\arrow["{\kappa_{10}}", from=3-3, to=3-4]
			\arrow["{\kappa_{11}}", shift left, from=3-4, to=3-5]
			\arrow["{\kappa_{12}}", shift left, from=3-5, to=3-4]
		\end{tikzcd}
	\end{equation}
	
	\begin{description}
		\item[Step 1:] The first step is to write out the corresponding ODE system:
		\begin{subequations} \label{eg:2 site ODEs}
			\begin{align} 
				\dot{s_0}  &=  -k_1s_0e+k_{11}y_4-k_{12}s_0f, \\
				\dot{s_1}  &=  k_2y_1-(k_3+k_4)s_1e +k_8y_3 -(k_9+k_{10})s_1f , \\
				\dot{s_2}   &=  k_5y_2 -k_6s_2e -k_7s_2f    , \\
				\dot{y_1}   &=  k_1s_0e-k_2y_1 + k_3s_1e , \\
				\dot{y_2}   &=  k_4s_1e-k_5y_2 +k_6s_2e     , \\
				\dot{y_3}   &=  k_7s_2f -k_8y_3+k_9s_1f    , \\
				\dot{y_4}   &=  k_{10}s_1f -k_{11}y_4 +k_{12}s_0f    , \\
				\dot{e}  &=  -k_1s_0e +k_2y_1 -(k_3+k_4)s_1e +k_5y_2 -k_6s_2e,     \\
				\dot{f}  &=  -k_7s_2f +k_8y_3 -(k_9+k_{10})s_1f +k_{11}y_4 -k_{12}s_0f.  
			\end{align}
		\end{subequations}
		
		The stoiciometric matrix is
		\begin{equation} \label{eg:2 site stoichiometric matrix}
			N = 
			\left[
			\begin{array}{*{12}c}
				-1 & 0 & 0 & 0 & 0 & 0 & 0 & 0 & 0 & 0 & 1 & -1 \\
				0 & 1 & -1 & -1 & 0 & 0 & 0 & 1 & -1 & -1 & 0 & 0 \\
				0 & 0 & 0 & 0 & 1 & -1 & -1 & 0 & 0 & 0 & 0 & 0 \\
				1 & -1 & 1 & 0 & 0 & 0 & 0 & 0 & 0 & 0 & 0 & 0 \\
				0 & 0 & 0 & 1 & -1 & 1 & 0 & 0 & 0 & 0 & 0 & 0 \\
				0 & 0 & 0 & 0 & 0 & 0 & 1 & -1 & 1 & 0 & 0 & 0 \\
				0 & 0 & 0 & 0 & 0 & 0 & 0 & 0 & 0 & 1 & -1 & 1 \\
				-1 & 1 & -1 & -1 & 1 & -1 & 0 & 0 & 0 & 0 & 0 & 0 \\
				0 & 0 & 0 & 0 & 0 & 0 & -1 & 1 & -1 & -1 & 1 & -1 
			\end{array}
			\right],
		\end{equation}		
		which has $\rank(N) = 6$. We conclude that $\rank(W) = 3$ and can see that a row-reduced $W$ is
		\begin{equation} \label{eg:2 site conservation matrix}
			W = 
			\left[
			\begin{array}{*{9}c}
				1 & 1 & 1 & 1 & 1 & 1 & 1 & 0 & 0 \\
				0 & 0 & 0 & 1 & 1 & 0 & 0 & 1 & 0 \\
				0 & 0 & 0 & 0 & 0 & 1 & 1 & 0 & 1 
			\end{array}
			\right].
		\end{equation}
		\item[Step 2:] We can easily and quickly see that by adding each row of $W$, we get a strictly positive vector $v = (1,1,1,2,2,2,2,1,1)$ such that $v \cdot N = 0$, since $WN = 0$ by construction. Thus our network is conservative which implies it is dissipative.
		
		\item[Step 3:] Next we look at the set of siphons for our network:
		\begin{subequations}    
			\begin{align} \label{eq:2site:siphons}
				Z_1 &= \{S_0,S_1,S_2,Y_1,Y_2,Y_3,Y_4 \}, \\
				Z_2 &= \{Y_1,Y_2,E \}, \label{eq:2site:siphon:z2} \\ 
				Z_3 &= \{Y_3,Y_4,F \}, \label{eq:2site:siphon:z3}\\
				Z_4 &= \{S_0,S_1,S_2,Y_1,Y_2,Y_3,Y_4, E, F \}, \\ 
				Z_5 &= \{S_0, Y_3,Y_4,F \}, \\
				Z_6 &= \{S_0,S_1,S_2,Y_1,Y_2,Y_3,Y_4, F \}, \\ 
				Z_7 &= \{S_0,S_1,S_2,Y_1,Y_2,Y_3,Y_4, E \}, \\
				Z_8 &= \{S_0,S_1, Y_4, Y_1, Y_2, E \}.  
			\end{align}
		\end{subequations}
		
		Note that Equation \eqref{eq:2site:siphon:z2} and Equation \eqref{eq:2site:siphon:z3} are the minimal siphons for this network and both correspond to conservation laws. We conclude that there are no relevant boundary equilibria.
		
		\item[Step 4:] We start by setting $c = (c_1,c_2,c_3) \in \mathbb{R}^3_{\geq 0}$ to get
		
		\begin{equation}
			\phi_c(x) = \begin{pmatrix}
				s_0 + s_1 + s_2 + y_1 + y_2 +y_3 +y_4 - c_1 \\
				k_2y_1-(k_3+k_4)s_1e +k_8y_3 -(k_9+k_{10})s_1f- \\
				k_5y_2 -k_6s_2e -k_7s_2f \\
				y_1 +y_2 +e -c_2 \\
				k_4s_1e-k_5y_2 +k_6s_2e  \\
				y_3+y_4+f \\
				-k_1s_0e +k_2y_1 -(k_3+k_4)s_1e +k_5y_2 -k_6s_2e          \\
				-k_7s_2f +k_8y_3 -(k_9+k_{10})s_1f +k_{11}y_4 -k_{12}s_0f
			\end{pmatrix}.
		\end{equation}		
		Taking the Jacobian gives us  		
		\begin{multline} \label{eq:2 site M}
			M(x) = 
			\left[
			\begin{array}{*{6}c}
				1 & 1 & 1 & 1 & 1 & 1   \\
				0 & -(k_3+k_4)e -(k_9+k_{10})f & 0 & k_2 & 0 & k_8   \\
				0 & 0 & -k_6e -k_7f & 0 & k_5 & 0  \\
				0 & 0 & 0 & 1 & 1 & 0   \\
				0 & k_4e & k_6e & 0 & -k_5 & 0   \\
				0 & 0 & 0 & 0 & 0 & 1    \\
				k_{12}f & k_{10}f & 0 & 0 & 0 & 0   \\
				-k_1e & -(k_3+k_4)e & -k_6e & k_2 & k_5 & 0   \\
				-k_{12}f & -(k_9+k_{10})f & -k_7f & 0 & 0 & k_8   \\
			\end{array} 
			\right. \\
				\left. 
			\begin{array}{*{9}c}  
				1 &  0 & 0 \\
				0 & -(k_3+k_4)s_1 & -(k_9+k_{10})s_1 \\
				0 & -k_6s_2 & -k_7s_2 \\
				0 & 1 & 0 \\
				0 & k_4s_1 + k_6s_2 & 0 \\
				1 &  0 & 1  \\
				-k_{11} & 0 & k_{10}s_1 + k_{12}s_0 \\
				0 & -k_1s_0-(k_3+k_4)s_1-k_6s_2 & 0 \\
				k_{11} & 0 & -k_{12}s_0 -(k_9 + k_{10})s_1 -k_7s_2
			\end{array}
			\right].
		\end{multline}		
		
		\item[Step 5:] Next we look at $\det(M(x))$:
		
		\begin{multline*}
			\det(M(x)) = e^{2} k_1 k_{1 1} k_2 k_4 k_5 k_8 + e^{3} k_1 k_{1 1} k_2 k_4 k_6 k_8 + f^{2} k_{1 0} k_{1 1} k_2 k_5 k_7 k_8 \\
			+ f^{3} k_{1 0} k_{1 2} k_2 k_5 k_7 k_8  + e f k_1 k_{1 1} k_2 k_5 k_7 k_8 \\
			+ f^{2} e k_1 k_{1 1} k_2 k_5 k_7 k_9 + e^{2} f k_1 k_{1 1} k_2 k_4 k_5 k_7 \\
			+ e^{2} f k_1 k_{1 1} k_2 k_4 k_7 k_8 + e^{2} f k_1 k_{1 1} k_3 k_5 k_7 k_8 \\
			+ e^{2} k_1 k_{1 2} k_2 k_4 k_5 k_8 s_0 + f^{2} k_1 k_{1 0} k_{1 1} k_5 k_7 k_8 s_0 \\
			+ e^{3} k_1 k_{1 1} k_2 k_4 k_6 k_7 s_2 + s_1^{2} e^{2} k_1 k_{1 0} k_2 k_4 k_6 k_8 \\
			+ f^{3} k_1 k_{1 0} k_{1 2} k_5 k_7 k_8 s_0 + e^{2} k_1 k_{1 1} k_2 k_4 k_6 k_8 s_0 \\
			+ f^{2} e k_1 k_{1 0} k_{1 1} k_5 k_7 k_8 + e^{2} k_1 k_{1 1} k_2 k_4 k_5 k_7 s_2 \\
			+ e^{2} k_1 k_{1 0} k_2 k_4 k_5 k_8 s_1 + e^{2} k_1 k_{1 1} k_2 k_4 k_6 k_8 s_1 \\
			+ e^{2} k_1 k_{1 1} k_2 k_4 k_6 k_8 s_2 + e^{2} k_1 k_{1 1} k_2 k_4 k_5 k_9 s_1 \\
			+ f^{2} e k_1 k_{1 0} k_2 k_5 k_7 k_8 + s_2^{2} e^{2} k_1 k_{1 1} k_2 k_4 k_6 k_7 \\
			+ s_0^{2} f^{2} k_1 k_{1 0} k_{1 2} k_5 k_7 k_8 + e^{3} k_1 k_{1 2} k_2 k_4 k_6 k_8 s_0 \\
			+ f^{2} k_{1 0} k_{1 2} k_2 k_5 k_7 k_8 s_0 + e^{3} k_1 k_{1 0} k_2 k_4 k_6 k_8 s_1 \\
			+ e^{3} k_1 k_{1 1} k_2 k_4 k_6 k_9 s_1 + f^{2} k_{1 0} k_{1 1} k_2 k_4 k_7 k_8 s_1 \\
			+ f^{2} k_{1 0} k_{1 1} k_2 k_6 k_7 k_8 s_2 + f^{2} k_{1 0} k_{1 2} k_2 k_5 k_7 k_8 s_1 
		\end{multline*}		
		\begin{multline*}
			+ f^{2} k_{1 0} k_{1 2} k_2 k_5 k_7 k_8 s_2 + s_0^{2} e^{2} k_1 k_{1 2} k_2 k_4 k_6 k_8 + s_1^{2} e^{2} k_1 k_{1 1} k_2 k_4 k_6 k_9 \\
			+ f^{2} k_{1 0} k_{1 1} k_3 k_5 k_7 k_8 s_1 + f^{3} k_{1 0} k_{1 2} k_2 k_4 k_7 k_8 s_1 + s_1^{2} f^{2} k_{1 0} k_{1 2} k_2 k_4 k_7 k_8 \\
			+ f^{3} k_{1 0} k_{1 2} k_2 k_6 k_7 k_8 s_2 + s_2^{2} f^{2} k_{1 0} k_{1 2} k_2 k_6 k_7 k_8 + f^{3} k_{1 0} k_{1 2} k_3 k_5 k_7 k_8 s_1 \\
			+ s_1^{2} f^{2} k_{1 0} k_{1 2} k_3 k_5 k_7 k_8 + e f k_1 k_{1 1} k_2 k_4 k_7 k_8 s_0  + e f k_1 k_{1 1} k_2 k_4 k_7 k_8 s_1 \\
			+ e f k_1 k_{1 1} k_3 k_5 k_7 k_8 s_0 + e f k_1 k_{1 0} k_2 k_5 k_7 k_8 s_1 + e f k_1 k_{1 1} k_3 k_5 k_7 k_8 s_1 \\
			+ e f k_1 k_{1 1} k_2 k_5 k_7 k_9 s_1 + e f k_{1 0} k_{1 1} k_2 k_4 k_6 k_8 s_1 + e f k_1 k_{1 2} k_2 k_5 k_7 k_8 s_0 \\
			+ s_1^{2} e f k_1 k_{1 0} k_2 k_4 k_7 k_8 + e f k_{1 0} k_{1 2} k_2 k_4 k_5 k_8 s_1 + s_1^{2} e f k_{1 0} k_{1 2} k_2 k_4 k_6 k_8 \\
			+ e f k_1 k_{1 0} k_2 k_5 k_7 k_8 s_2 + e f k_1 k_{1 1} k_2 k_5 k_7 k_9 s_2 + e f k_1 k_{1 1} k_2 k_6 k_7 k_8 s_2 \\
			+ f^{2} e k_1 k_{1 0} k_{1 2} k_5 k_7 k_8 s_0 + f^{2} e k_1 k_{1 0} k_2 k_4 k_7 k_8 s_1 + f^{2} e k_1 k_{1 0} k_3 k_5 k_7 k_8 s_1 \\
			+ f^{2} e k_1 k_{1 1} k_2 k_6 k_7 k_9 s_2 + e^{2} f k_1 k_{1 2} k_2 k_4 k_6 k_8 s_0 + f^{2} e k_{1 0} k_{1 2} k_2 k_4 k_7 k_8 s_1 \\
			+ f^{2} e k_1 k_{1 2} k_3 k_5 k_7 k_8 s_0 + e^{2} f k_1 k_{1 0} k_2 k_4 k_6 k_8 s_1 + e^{2} f k_1 k_{1 0} k_2 k_6 k_7 k_8 s_2 \\
			+ e^{2} f k_1 k_{1 1} k_2 k_4 k_6 k_9 s_1 + e^{2} f k_1 k_{1 1} k_2 k_4 k_7 k_9 s_1 + e^{2} f k_1 k_{1 0} k_3 k_5 k_7 k_8 s_1 \\
			+ e^{2} f k_{1 0} k_{1 2} k_2 k_4 k_6 k_8 s_1 + e^{2} f k_1 k_{1 1} k_2 k_6 k_7 k_9 s_2 + f^{2} e k_1 k_{1 0} k_2 k_6 k_7 k_8 s_2 \\
			+ e^{2} f k_1 k_{1 1} k_3 k_5 k_7 k_9 s_1 + f^{2} e k_{1 0} k_{1 2} k_2 k_6 k_7 k_8 s_2 + e^{2} f k_1 k_{1 2} k_2 k_4 k_7 k_8 s_0 \\
			+ e^{2} f k_1 k_{1 2} k_3 k_5 k_7 k_8 s_0 + e^{2} k_1 k_{1 1} k_2 k_4 k_6 k_7 s_1 s_2 + e^{2} k_1 k_{1 0} k_2 k_4 k_6 k_8 s_0 s_1 \\
			+ e^{2} k_1 k_{1 0} k_2 k_4 k_6 k_8 s_1 s_2 + e^{2} k_1 k_{1 1} k_2 k_4 k_6 k_9 s_1 s_2 + e^{2} k_1 k_{1 2} k_2 k_4 k_6 k_8 s_0 s_1 \\
			+ e^{2} k_1 k_{1 2} k_2 k_4 k_6 k_8 s_0 s_2 + s_0^{2} e f k_1 k_{1 2} k_2 k_4 k_7 k_8 + s_1^{2} e f k_1 k_{1 0} k_3 k_5 k_7 k_8 \\
			+ s_1^{2} e f k_1 k_{1 1} k_3 k_5 k_7 k_9 + f^{2} e k_{1 0} k_{1 2} k_2 k_4 k_6 k_8 s_1 + e^{2} k_1 k_{1 1} k_2 k_4 k_6 k_7 s_0 s_2 \\
			+ f^{2} e k_1 k_{1 2} k_2 k_4 k_7 k_8 s_0 + f^{2} e k_1 k_{1 1} k_2 k_4 k_7 k_9 s_1 + e^{2} k_1 k_{1 1} k_2 k_4 k_6 k_9 s_0 s_1 \\
			+ s_1^{2} e f k_1 k_{1 1} k_2 k_4 k_7 k_9 + f^{2} e k_1 k_{1 1} k_3 k_5 k_7 k_9 s_1 + f^{2} k_1 k_{1 0} k_{1 2} k_5 k_7 k_8 s_0 s_1 \\
			+ f^{2} k_1 k_{1 0} k_{1 2} k_5 k_7 k_8 s_0 s_2 + f^{2} k_{1 0} k_{1 2} k_2 k_4 k_7 k_8 s_0 s_1 + f^{2} k_{1 0} k_{1 2} k_3 k_5 k_7 k_8 s_0 s_1 \\
			+ f^{2} k_{1 0} k_{1 2} k_2 k_4 k_7 k_8 s_1 s_2 + f^{2} k_{1 0} k_{1 2} k_2 k_6 k_7 k_8 s_0 s_2 + f^{2} k_{1 0} k_{1 2} k_2 k_6 k_7 k_8 s_1 s_2 \\
			+ f^{2} k_{1 0} k_{1 2} k_3 k_5 k_7 k_8 s_1 s_2 + s_0^{2} e f k_1 k_{1 2} k_3 k_5 k_7 k_8 + s_2^{2} e f k_1 k_{1 0} k_2 k_6 k_7 k_8 \\
			+ f^{2} e k_{1 0} k_{1 2} k_3 k_5 k_7 k_8 s_1 + s_2^{2} e f k_1 k_{1 1} k_2 k_6 k_7 k_9 + e^{2} f k_1 k_{1 1} k_2 k_4 k_6 k_7 s_2 \\
			+ e^{2} f k_1 k_{1 0} k_2 k_4 k_7 k_8 s_1 - e f k_1 k_{1 0} k_{1 1} k_4 k_5 k_8 s_1 - e f k_{1 0} k_{1 1} k_2 k_4 k_5 k_7 s_1 \\
			+ e f k_1 k_{1 0} k_2 k_4 k_7 k_8 s_0 s_1 + e f k_1 k_{1 2} k_2 k_4 k_7 k_8 s_0 s_2 + e f k_1 k_{1 0} k_3 k_5 k_7 k_8 s_0 s_1 \\
			+ e f k_1 k_{1 0} k_2 k_4 k_7 k_8 s_1 s_2 + e f k_1 k_{1 2} k_2 k_4 k_7 k_8 s_0 s_1 - s_1^{2} e f k_1 k_{1 0} k_{1 1} k_4 k_5 k_7 \\
		\end{multline*}		
		\begin{multline}
			+ e f k_1 k_{1 2} k_3 k_5 k_7 k_8 s_0 s_1 - e^{2} f k_1 k_{1 0} k_{1 1} k_4 k_5 k_7 s_1 + e f k_1 k_{1 2} k_3 k_5 k_7 k_8 s_0 s_2 \\
			+ e f k_{1 0} k_{1 2} k_2 k_4 k_6 k_8 s_0 s_1 + e f k_{1 0} k_{1 2} k_2 k_4 k_6 k_8 s_1 s_2 - f^{2} e k_1 k_{1 0} k_{1 1} k_4 k_5 k_7 s_1 \\
			+ e f k_1 k_{1 0} k_2 k_6 k_7 k_8 s_0 s_2 + e f k_1 k_{1 0} k_2 k_6 k_7 k_8 s_1 s_2 + e f k_1 k_{1 1} k_2 k_4 k_7 k_9 s_0 s_1 \\
			+ e f k_1 k_{1 1} k_3 k_5 k_7 k_9 s_0 s_1 + e f k_1 k_{1 1} k_2 k_4 k_7 k_9 s_1 s_2 + e f k_1 k_{1 1} k_2 k_6 k_7 k_9 s_0 s_2 \\
			+ e f k_1 k_{1 1} k_2 k_6 k_7 k_9 s_1 s_2 + e f k_1 k_{1 0} k_3 k_5 k_7 k_8 s_1 s_2 + e f k_1 k_{1 1} k_3 k_5 k_7 k_9 s_1 s_2 \\
			- e f k_1 k_{1 0} k_{1 1} k_4 k_5 k_7 s_0 s_1 - e f k_1 k_{1 0} k_{1 1} k_4 k_5 k_7 s_1 s_2. 
		\end{multline}
		
		From here we notice that we have mostly positive terms but some are negative. Indeed $(-1)^s = (-1)^6 =1$, so if all terms were positive we would conclude that this reaction network is monostationary for any assignment of rate parameters and in all relevant stoichiometric compatability classes. This is not the case however, so we move on to Step 6.
		
		\item[Step 6:] We note that the ODEs for $s_1,e,f$ contain the most terms. Solving those ODEs at steady state for their respective species at steady state will involve the most terms. So we will start by letting $s_1,e,f$ be free variables. We need to solve for $s_0,s_2,y_1,y_2,y_3,y_4$ in terms of just $s_1,e,f$ and our parameters. Setting the derivatives to $0$ in each equation of the system defined in Equations \eqref{eg:2 site ODEs} gives		
		\begin{subequations} 
			\begin{align} 
				0  &=  -k_1s_0e+k_{11}y_4-k_{12}s_0f \label{eg:2 site ODEs at steady state s0},\\
				0 &= k_2y_1-(k_3+k_4)s_1e +k_8y_3 -(k_9+k_{10})s_1f,  \label{eg:2 site ODEs at steady state s1}\\
				0   &=  k_5y_2 -k_6s_2e -k_7s_2f,    \label{eg:2 site ODEs at steady state s2} \\
				0   &=  k_1s_0e-k_2y_1 + k_3s_1e,  \label{eg:2 site ODEs at steady state y1}\\
				0   &= k_4s_1e-k_5y_2 +k_6s_2e,     \label{eg:2 site ODEs at steady state y2} \\
				0  &=  k_7s_2f -k_8y_3+k_9s_1f,   \label{eg:2 site ODEs at steady state y3}  \\
				0   &=  k_{10}s_1f -k_{11}y_4 +k_{12}s_0f,    \label{eg:2 site ODEs at steady state y4} \\
				0  &= -k_1s_0e +k_2y_1 -(k_3+k_4)s_1e +k_5y_2 -k_6s_2e,  \label{eg:2 site ODEs at steady state e}  \\
				0   &=  -k_7s_2f +k_8y_3 -(k_9+k_{10})s_1f +k_{11}y_4 -k_{12}s_0f. \label{eg:2 site ODEs at steady state f} 
			\end{align}
		\end{subequations}
		
		Adding Equation \eqref{eg:2 site ODEs at steady state s0} and Equation \eqref{eg:2 site ODEs at steady state y4}, we can solve for $s_0$; $0 = -k_1s_0e +k_{10}s_1f$ implies
		\begin{equation}
			s_0 = \frac{k_{10}s_1 f}{k_1e}. \label{eg:2 site positive parametrization s0}
		\end{equation}		
		Similarly we can add Equation $\eqref{eg:2 site ODEs at steady state s2}$ and Equation \eqref{eg:2 site ODEs at steady state y2} and solve for $s_2$; $0 =k_4s_1e -k_7s_2f$ implies
		\begin{equation}
			s_2 = \frac{k_4s_1e}{k_7f}. \label{eg:2 site positive parametrization s2}
		\end{equation}		
		Then we can solve for $y_1$ through $y_4$ in their respective equations and plug in Equations \eqref{eg:2 site positive parametrization s0} and \eqref{eg:2 site positive parametrization s2} as needed to obtain the remaining expressions in terms of $s_1,e,f$:
		\begin{subequations}
		\begin{align}
			y_1 &= \frac{k_1s_0e + k_3s_1e}{k_2}= \frac{k_1e(\frac{k_{10}s_1f}{k_1e}) + k_3s_1}{k_2} \nonumber \\
			&= \frac{k_{10}s_1f + k_3s_1e}{k_2} \label{eg:2 site positive parametrization y1} \\
			y_2 &= \frac{k_4s_1e +k_6s_2e}{k_5} = \frac{k_4s_1e + k_6e(\frac{k_4s_1e}{k_7f})}{k_5} \nonumber \\
			&= \frac{k_4k_7s_1ef + k_4k_6s_1e^2}{k_5k_7f} \label{eg:2 site positive parametrization y2}\\
			y_3 &= \frac{k_7s_2f +k_9s_1f}{k_8} = \frac{k_7f(\frac{k_4s_1e}{k_7f}) +k_9s_1f}{k_8} \nonumber \\
			&= \frac{k_4s_1e + k_9s_1f}{k_8} \label{eg:2 site positive parametrization y3} \\
			y_4 &= \frac{k_{10}s_1f + k_{12}s_0f}{k_{11}} = \frac{k_{10}s_1f + k_{12}f(\frac{k_{10}s_1f}{k_1e})}{k_{11}} \nonumber \\
			&= \frac{k_1k_{10}s_1ef + k_{10}k_{12}s_1f^2}{k_1k_{11}e} \label{eg:2 site positive parametrization y4}
		\end{align}
		\end{subequations}
		
Thus $\Tilde{x} = (s_1,e,f)$ and we have found a positive parametrization $\psi(\tilde{x})$:
		\begin{multline} \label{eq:psi 2site phospho}
			\psi(s_1,e,f) = \left(\frac{k_{10}s_1f}{k_1e}, s_1, \frac{k_4s_1e}{k_7f},\frac{k_{10}s_1f + k_3s_1e}{k_2},\frac{k_4k_7s_1ef + k_4k_6s_1e^2}{k_5k_7f}, \right. \\
			\left. \frac{k_4s_1e + k_9s_1f}{k_8},  \frac{k_1k_{10}s_1ef + k_{10}k_{12}s_1f^2}{k_1k_{11}e}, e, f \right). 
		\end{multline}
		Now we just need to plug $\psi$ into our matrix $M(x)$, defined in Equation \eqref{eq:2 site M}, to get 		
		\begin{multline} \label{eg:2 site M(psi(x))} 
			M(\psi(\tilde{x})) =
			\left[
			\begin{array}{*{6}c}
				1 & 1 & 1 & 1 & 1 & 1   \\
				0 & -(k_3+k_4)e -(k_9+k_{10})f & 0 & k_2 & 0 & k_8   \\
				0 & 0 & -k_6e -k_7f & 0 & k_5 & 0   \\
				0 & 0 & 0 & 1 & 1 & 0   \\
				0 & k_4e & k_6e & 0 & -k_5 & 0   \\
				0 & 0 & 0 & 0 & 0 & 1    \\
				k_{12}f & k_{10}f & 0 & 0 & 0 & 0   \\
				-k_1e & -(k_3+k_4)e & -k_6e & k_2 & k_5 & 0   \\
				-k_{12}f & -(k_9+k_{10})f & -k_7f & 0 & 0 & k_8 \\
			\end{array} 
			\right. \\ \left. \footnotesize
			\begin{array}{*{9}c}
				1 & 0 & 0 \\
				0 & -(k_3+k_4)s_1 & -(k_9+k_{10})s_1 \\
				0 &  -\frac{k_4k_6s_1e}{k_7f} & -\frac{k_4s_1e}{f} \\
				0 & 1 & 0 \\
				0 & k_4s_1 + \frac{k_4k_6s_1e}{k_7f} & 0 \\
				1 & 0 & 1  \\
				-k_{11} & 0 & k_{10}s_1 + \frac{k_{10}k_{12}s_1f}{k_1e} \\
				0 & -\frac{k_{10}s_1f}{e}-(k_3+k_4)s_1-\frac{k_4k_6s_1e}{k_7f} & 0 \\
				k_{11} & 0 & -\frac{k_{10}k_{12}s_1f}{k_1e} -(k_9 + k_{10})s_1 -\frac{k_4s_1e}{f} 
			\end{array}
			\right].
		\end{multline}
		
		We can now take the determinant of this matrix. It is a large rational expression in terms of the parameters $\{k_i\}$  and species $s_1,e,f$.  We break this expression up into a numerator $\text{num}(\det(M(\psi(X))))$ and denominator $\text{den}(\det(M(\psi(X))))$ to get		
		\begin{equation}
			\text{den}(\det(M(\psi(X)))) =k_1k_7e^2f^2. \label{eq:den(M(psi(x)))}
		\end{equation}
		
		We note that $\text{den}(\det(M(\psi(X))))$ is positive for any assignment of $\{k_i\}$ and any values of $s_1,e,f$. Thus the sign of $\det(M(\psi(X)))$ depends only on the monomials in the numerator, 		
		\begin{multline*} 
			\text{num}(\det(M(\psi(X)))) = \\
			s_1^{2} k_4^{3} k_1^{2} e^{6} k_{1 1} k_2 k_6 + s_1^{2} k_7^{2} k_{1 0}^{3} f^{6} k_{1 2} k_5 k_8 + k_7^{2} k_1^{2} f^{3} e^{3} k_{1 1} k_2 k_5 k_8 \\
			+ k_7^{2} k_1^{2} f^{3} e^{4} k_{1 1} k_3 k_5 k_8 + k_7^{2} k_1^{2} f^{4} e^{3} k_{1 0} k_{1 1} k_5 k_8 + k_7^{2} k_1^{2} f^{4} e^{3} k_{1 0} k_2 k_5 k_8 \\
			+ k_7^{2} k_1^{2} f^{4} e^{3} k_{1 1} k_2 k_5 k_9 + k_7^{2} k_1^{2} f^{3} e^{4} k_{1 1} k_2 k_4 k_5 + k_7^{2} k_1^{2} f^{3} e^{4} k_{1 1} k_2 k_4 k_8 \\
			+ k_1^{2} f^{2} e^{4} k_{1 1} k_2 k_4 k_5 k_7 k_8 + k_4^{2} k_1^{2} e^{5} f k_{1 1} k_2 k_5 k_7 s_1 + k_4^{2} k_1^{2} e^{5} f k_{1 1} k_2 k_6 k_8 s_1 \\
			+ k_4^{2} k_1^{2} e^{6} f k_{1 1} k_2 k_6 k_7 s_1 + k_1^{2} f^{2} e^{5} k_{1 1} k_2 k_4 k_6 k_7 k_8 + k_7^{2} f^{4} e^{2} k_1 k_{1 0} k_{1 1} k_2 k_5 k_8 \\
			+ k_7^{2} f^{5} e^{2} k_1 k_{1 0} k_{1 2} k_2 k_5 k_8 + k_7^{2} k_1^{2} f^{3} e^{3} k_{1 1} k_2 k_4 k_8 s_1 + k_7^{2} k_1^{2} f^{3} e^{3} k_{1 0} k_2 k_5 k_8 s_1 \\
			+ k_7^{2} k_1^{2} f^{3} e^{3} k_{1 1} k_2 k_5 k_9 s_1 + k_7^{2} k_{1 0}^{2} f^{5} e^{2} k_1 k_{1 2} k_5 k_8 s_1 + s_1^{2} k_7^{2} k_{1 0}^{2} f^{4} e^{2} k_1 k_2 k_4 k_8 \\
			+ k_7^{2} k_1^{2} f^{4} e^{3} k_{1 0} k_2 k_4 k_8 s_1 + k_7^{2} k_1^{2} f^{3} e^{4} k_{1 0} k_2 k_4 k_8 s_1 + s_1^{2} k_7^{2} k_1^{2} f^{3} e^{3} k_{1 0} k_3 k_5 k_8 \\
			+ k_7^{2} k_1^{2} f^{3} e^{4} k_{1 1} k_2 k_4 k_9 s_1 + k_7^{2} k_{1 0}^{2} f^{5} e k_1 k_{1 1} k_5 k_8 s_1 + k_7^{2} k_1^{2} f^{3} e^{4} k_{1 0} k_3 k_5 k_8 s_1 \\
			+ k_7^{2} k_1^{2} f^{4} e^{3} k_{1 0} k_3 k_5 k_8 s_1 + k_7^{2} k_{1 0}^{2} f^{5} e k_{1 2} k_2 k_5 k_8 s_1 + k_4^{2} k_1^{2} f^{2} e^{5} k_{1 1} k_2 k_6 k_7 s_1 \\
			+ k_7^{2} k_1^{2} f^{4} e^{3} k_{1 1} k_2 k_4 k_9 s_1 + k_7^{2} k_{1 0}^{2} f^{6} e k_1 k_{1 2} k_5 k_8 s_1 + k_7^{2} k_1^{2} f^{3} e^{3} k_{1 1} k_3 k_5 k_8 s_1 \\
			+ k_7^{2} k_1^{2} f^{4} e^{3} k_{1 1} k_3 k_5 k_9 s_1 + k_7^{2} k_1^{2} f^{3} e^{4} k_{1 1} k_3 k_5 k_9 s_1 + s_1^{2} k_4^{2} k_1^{2} e^{5} f k_{1 1} k_2 k_6 k_7 \\
			+ s_1^{2} k_7^{2} k_{1 0}^{2} f^{4} e^{2} k_1 k_3 k_5 k_8 + s_1^{2} k_7^{2} k_1^{2} f^{3} e^{3} k_{1 0} k_2 k_4 k_8 + s_1^{2} k_7^{2} k_1^{2} f^{3} e^{3} k_{1 1} k_2 k_4 k_9 \\
			+ s_1^{2} k_7^{2} k_1^{2} f^{3} e^{3} k_{1 1} k_3 k_5 k_9 + s_1^{2} k_4^{2} k_1^{2} f^{2} e^{4} k_{1 0} k_2 k_7 k_8 + s_1^{2} k_7^{2} k_{1 0}^{2} f^{5} e k_1 k_{1 2} k_5 k_8 \\
			+ s_1^{2} k_4^{2} k_1^{2} f^{2} e^{4} k_{1 1} k_2 k_7 k_9 + s_1^{2} k_{1 0}^{2} f^{4} e^{2} k_1 k_{1 2} k_4 k_5 k_7 k_8 \\
			+ 2 s_1^{2} k_4^{2} k_1^{2} e^{5} f k_{1 0} k_2 k_6 k_8 + s_1^{2} k_7^{2} f^{4} e^{2} k_1 k_{1 0} k_{1 1} k_2 k_4 k_9 \\
			+ s_1^{2} k_7^{2} f^{4} e^{2} k_1 k_{1 0} k_{1 1} k_3 k_5 k_9 + s_1^{2} k_1^{2} f^{2} e^{4} k_{1 0} k_3 k_4 k_5 k_7 k_8 \\
			- k_7^{2} k_1^{2} f^{3} e^{4} k_{1 0} k_{1 1} k_4 k_5 s_1 - k_7^{2} k_1^{2} f^{4} e^{3} k_{1 0} k_{1 1} k_4 k_5 s_1 \\
			- s_1^{2} k_7^{2} k_1^{2} f^{3} e^{3} k_{1 0} k_{1 1} k_4 k_5 + 2 s_1^{2} k_7^{2} k_{1 0}^{2} f^{5} e k_{1 2} k_3 k_5 k_8 \\
			+ 2 s_1^{2} k_4^{2} k_1^{2} e^{5} f k_{1 1} k_2 k_6 k_9 + s_1^{2} k_1^{2} f^{2} e^{4} k_{1 1} k_3 k_4 k_5 k_7 k_9 \\
			- s_1^{2} k_4^{2} k_1^{2} f^{2} e^{4} k_{1 0} k_{1 1} k_5 k_7 + s_1^{2} k_4^{2} f^{2} e^{4} k_1 k_{1 0} k_{1 1} k_2 k_6 k_7 \\
			+ 2 s_1^{2} k_7^{2} k_{1 0}^{2} f^{5} e k_{1 2} k_2 k_4 k_8 - s_1^{2} k_7^{2} k_{1 0}^{2} f^{4} e^{2} k_1 k_{1 1} k_4 k_5 \\
			+ 2 k_7^{2} f^{5} e^{2} k_1 k_{1 0} k_{1 2} k_2 k_4 k_8 s_1 - k_7^{2} f^{3} e^{3} k_1 k_{1 0} k_{1 1} k_2 k_4 k_5 s_1 \\
		\end{multline*}		
		\begin{multline} \label{eg:2site num(M(psi(x)))}
			+ 2 k_7^{2} f^{4} e^{3} k_1 k_{1 0} k_{1 2} k_2 k_4 k_8 s_1 + 3 s_1^{2} k_4^{2} f^{2} e^{4} k_1 k_{1 0} k_{1 2} k_2 k_6 k_8 \\
			+ 2 s_1^{2} k_7^{2} f^{4} e^{2} k_1 k_{1 0} k_{1 2} k_2 k_4 k_8 + 2 s_1^{2} k_4^{2} f^{3} e^{3} k_1 k_{1 0} k_{1 2} k_2 k_7 k_8 \\
			+ 2 s_1^{2} k_{1 0}^{2} f^{3} e^{3} k_1 k_2 k_4 k_6 k_7 k_8 - k_1^{2} f^{3} e^{3} k_{1 0} k_{1 1} k_4 k_5 k_7 k_8 s_1 \\
			+ 2 k_1^{2} f^{2} e^{4} k_{1 0} k_2 k_4 k_5 k_7 k_8 s_1 + 2 k_1^{2} f^{2} e^{4} k_{1 1} k_2 k_4 k_5 k_7 k_9 s_1 \\
			+ 2 k_1^{2} f^{2} e^{5} k_{1 0} k_2 k_4 k_6 k_7 k_8 s_1 + 2 k_1^{2} f^{2} e^{4} k_{1 1} k_2 k_4 k_6 k_7 k_8 s_1 \\
			+ 2 k_1^{2} f^{3} e^{4} k_{1 0} k_2 k_4 k_6 k_7 k_8 s_1 + 2 k_1^{2} f^{3} e^{4} k_{1 1} k_2 k_4 k_6 k_7 k_9 s_1 \\
			+ 2 k_7^{2} f^{4} e^{2} k_1 k_{1 0} k_{1 1} k_2 k_4 k_8 s_1 + 2 k_7^{2} f^{4} e^{2} k_1 k_{1 0} k_{1 2} k_2 k_5 k_8 s_1 \\
			+ 2 k_7^{2} f^{4} e^{2} k_1 k_{1 0} k_{1 1} k_3 k_5 k_8 s_1 + 2 k_7^{2} f^{4} e^{3} k_1 k_{1 0} k_{1 2} k_3 k_5 k_8 s_1 \\
			+ 2 k_1^{2} f^{2} e^{5} k_{1 1} k_2 k_4 k_6 k_7 k_9 s_1 + 2 k_7^{2} f^{5} e^{2} k_1 k_{1 0} k_{1 2} k_3 k_5 k_8 s_1 \\
			+ 2 s_1^{2} k_1^{2} f^{2} e^{4} k_{1 0} k_2 k_4 k_6 k_7 k_8 + 2 s_1^{2} k_1^{2} f^{2} e^{4} k_{1 1} k_2 k_4 k_6 k_7 k_9 \\
			+ 2 s_1^{2} k_7^{2} f^{4} e^{2} k_1 k_{1 0} k_{1 2} k_3 k_5 k_8 + 3 s_1^{2} k_{1 0}^{2} f^{4} e^{2} k_{1 2} k_2 k_4 k_6 k_7 k_8 \\
			+ 3 f^{3} e^{3} k_1 k_{1 0} k_{1 1} k_2 k_4 k_6 k_7 k_8 s_1 + 2 s_1^{2} f^{3} e^{3} k_1 k_{1 0} k_{1 1} k_2 k_4 k_6 k_7 k_9 \\
			+ 2 s_1^{2} f^{3} e^{3} k_1 k_{1 0} k_{1 2} k_3 k_4 k_5 k_7 k_8 + 3 f^{3} e^{3} k_1 k_{1 0} k_{1 2} k_2 k_4 k_5 k_7 k_8 s_1 \\
			+ 3 f^{3} e^{4} k_1 k_{1 0} k_{1 2} k_2 k_4 k_6 k_7 k_8 s_1 + 3 f^{4} e^{3} k_1 k_{1 0} k_{1 2} k_2 k_4 k_6 k_7 k_8 s_1 \\
			+ 3 s_1^{2} f^{3} e^{3} k_1 k_{1 0} k_{1 2} k_2 k_4 k_6 k_7 k_8
		\end{multline}
		
		It is important to note that the construction of the Newton polytope depends entirely on the degree of the monomial terms and not the coefficients $\{k_i\}$, {\em i.e.,} to construct the Newton polytope there is no distinction between $a = s_1^2e^6$ and $b = k_1^2k_2k_4^2k_6k_{11}s_1^2e^6$ (they both belong to the same vertex in the Newton polytope). Using this, we can extract the 3-tuples in $\mathbb{N}^3$ corresponding to the multidegree of the monomials. The multidegree of $a$ and $b$ for example is $(2,6,0)$.
		
		We note that $\text{num}(\det(M(\psi(X))))$ is a polynomial with 25 distinct monomials, of which 6 have both $+$ and $-$ signs (see Table \ref{table:multidegrees}). 
		
		\begin{table} 
			\begin{center}
				\begin{tabular}{ |c|c|c|c|} 
					\hline
					(0,2,4) (+) & (1,1,5) (+) & (1,4,3) (+/-) & (2,3,3) (+/-) \\
					\hline
					(0,2,5) (+) & (1,1,6) (+) & (1,5,1) (+) & \textbf{(2,4,2) (+/-)} \\
					\hline
					(0,3,3) (+) & (1,2,4) (+) & (1,5,2) (+) & (2,5,1) (+) \\ 
					\hline
					(0,3,4) (+) & (1,2,5) (+) & (1,6,1) (+) & (2,6,0) (+) \\ 
					\hline
					(0,4,2) (+) & (1,3,3) (+/-) & (2,0,6) (+) &  \\ 
					\hline
					(0,4,3) (+) & (1,3,4) (+/-) & (2,1,5) (+) &  \\
					\hline
					(0,5,2) (+) & (1,4,2) (+) & (2,2,4) (+/-) &  \\
					\hline
				\end{tabular}
			\end{center}
			\caption{Multidegrees of the monomial terms in $\text{num}(\det(M(\psi(X))))$ and their signs (Equation \eqref{eg:2site num(M(psi(x)))}). Recall that the species corresponding to the sequences are of the form $(s_1,e,f)$. For example, the sequence (2,3,3) corresponds to any monomials of the form $\alpha s_1^2e^3f^3$ where $\alpha$ is some positive coefficient in terms of $k_1,...,k_{12}$.} \label{table:monomial multidegrees and their signs 2site phospho dephospho}  \label{table:multidegrees}
		\end{table}
		
		Now we look at the Newton polytope of $\text{num}(\det(M(\psi(X))))$. 
		
		\item[Step 7:]  We use the \texttt{Oscar.jl} \cite{OSCAR-book,OSCAR} package in Julia to construct the Newton polytope. We can define the monomials generated from the tuples in Table \ref{table:monomial multidegrees and their signs 2site phospho dephospho}. We define this polynomial as an element in the ring $\mathbb{Q}[s_1,e,f]$. Using \texttt{Oscar.jl}, we generate the associated Newton polytope and find the following vertices for the Newton polytope: (1,6,1), (1,1,6), (0,2,5), (2,4,2), (1,5,1), (0,4,2), (2,1,5), 0,5,2) and (0,2,4). 
		
		We notice that of the 6 monomials exhibiting sign changes, only (2,4,2) is a vertex of the Newton polytope. Upon further inspection we note that this monomial has 9 distinct coefficients, of which 8 are positive.  The coefficient terms for the (2,4,2) monomial are		
		\begin{multline} \label{eq:vertex (2,4,2) coefficients 2site phospho}
			-k_1^2k_4^2k_5k_7k_{10}k_{11}s_1^2e^4f^2 + k_1^2k_2k_4^2k_7k_8k_{10}s_1^2e^4f^2  \\
			+ 2k_1^2k_2k_4k_6k_7k_8k_{10}s_1^2e^4f^2 
			+ k_1^2k_3k_4k_5k_7k_8k_{10}s_1^2e^4f^2 \\ 
			+ k_1^2k_2k_4^2k_7k_9k_{11}s_1^2e^4f^2 + 2k_1^2k_2k_4k_6k_7k_9k_{11}s_1^2e^4f^2 \\
			+k_1^2k_3k_4k_5k_7k_9k_{11}s_1^2e^4f^2 + k_1k_2k_4^2k_6k_7k_{10}k_{11}s_1^2e^4f^2 \\
			+ 3k_1k_2k_4^2k_6k_8k_{10}k_{12}s_1^2e^4f^2.
		\end{multline}		
		We can factor out $k_1k_4s_1^2e^4f^2$ from each of the terms. Thus the monomial $s_1^2e^4f^2$ is negative if and only if 
		\begin{multline}\label{ineq:vertex (2,4,2) coefficients 2site phospho}
			k_1k_4k_5k_7k_{10}k_{11} > k_1k_2k_4k_7k_8k_{10} + 2k_1k_2k_6k_7k_8k_{10} + k_1k_3k_6k_7k_8k_{10}  \\ + k_1k_2k_4k_7k_9k_{11} 
			+ 2k_1k_2k_6k_7k_9k_{11} + k_1k_3k_5k_7k_9k_{11}  \\ + k_2k_4k_6k_7k_{10}k_{11} + 3k_2k_4k_6k_8k_{10}k_{12}.
		\end{multline}
		
		Thus there exists $(s_1,e,f) \in \mathbb{R}_+^3$ such that the inequality in Equation \eqref{ineq:vertex (2,4,2) coefficients 2site phospho} holds. Then $\text{sign}(\text{num}(\det(M(\psi(X))))) = -1$ and there exists some $c$ such that the $\mathcal{P}^+_c \neq \emptyset$ admits multiple steady states. We thus determine that the 2-site phosphorylation dephosphorylation cycle network is multistationary and that we can find multiple steady states in the parameter region defined by Equation $\eqref{ineq:vertex (2,4,2) coefficients 2site phospho}$.
		
Note that we checked multistationarity of a specification of rate constants $k_1,...,k_{12}$ that satisfy Equation \eqref{ineq:vertex (2,4,2) coefficients 2site phospho}. For this network we do indeed obtain values for two stable steady states of the system. 
		
		It should be noted that for the other five monomials with negative terms, it is not clear whether or not these terms can dominate the polynomial in Equation \eqref{eg:2site num(M(psi(x)))}. Because of this we do not know if we can fully characterize the regions of multistationarity for this reaction network. 
	\end{description}

	One could continue this procedure for values of $n>2$ in the n-site phosphorylation dephosphorylation cycle reaction network, but the computational costs grow very quickly. The network is characterized the same way as the network in Equation \eqref{RN:1 site phosphorylation dephosphorylation standard}, where reversibility is posited on the left of the intermediate species (ours is on the right). They have a number of similar characteristics as $n$ increases. First, the number of species grows by $3n+3$, adding in one extra substrate $S_n$ and two intermediates $Y_{2n+1}$ and $Y_{2n+2}$. Additionally, the number of reactions in the network is $6n$. It should also be noted that the rank of the network is always $3n$, and thus its corank (rank of the conservation matrix) is 3. Recall that the stoichiometric matrix $N$ has size $|\mathcal{S}| \times |\mathcal{R}|$, and thus its rank would be at most $|\mathcal{S}| = 3n+3$.  One can do a proof by induction on the rank of $N$ to show this.

	\subsection{Mass-action Chemostat Model} \label{section:chemostat}
	
	In this section we will define a mass-action reaction network based on a model of a subsystem of the human gut microbiota \cite{Adrian_Ayati_Mangalam_2023}.  The original network was based on a continuously-operated bioreactor called a chemostat \cite{Plazl_Znidarsic-Plazl_2019}. Chemostats are frequently used to gather data for idealized microbial ecologies \cite{WINDER2011261}. One question is if conclusions about the mass-action network can give us insights into the motivating chemostat-model network. This would allow modelers to work with the potentially simpler mass-action model and perhaps guide the analysis of the chemostat model. 
	
The ODE system in \cite{Adrian_Ayati_Mangalam_2023} involves many Monod forms to model the dynamics. A Monod form is a term of the form 
	$$\Psi(S; \gamma, \Psi_{\max}) = \Psi_{\max} \frac{S}{\gamma + S},$$
	where $S$ is the concentration of the limiting substrate, $\Psi_{\max}$ is the asymptotic value of $Psi(S)$ as $S$ increases (the rate constant at saturation), and $\gamma$ is the shape constant or the half-velocity constant, {\em i.e.,} the value of $S$ when $$\frac{\Psi(S; \gamma, \Psi_{\max})}{\Psi_{\max}} = \frac{1}{2}.$$  There is a Monod form for each of the microorganisms of interest, each with a possibly distinct limiting substrate $S$. 
	
	The methods in \cite{multistationarity_regions2017} are defined for reaction networks with mass-action kinetics, and not for Monod kinetics. By making some simplifying assumptions, we can write a surrogate model based on mass-action kinetics that result in polynomial equations.

Our network contains nodes for the microbial species \textit{Bacteroides thetaiotaomicron}, \textit{ Methanobrevibacter smithii}, and \textit{Eubacterium rectale}. The reaction network also includes a node for polysacharides, a node for carbon dioxide and hydrogen, and a node for acetate. We will use the same variables introduced in \cite{Adrian_Ayati_Mangalam_2023}:
	
	\begin{itemize} \label{def:project3:species}
		\item $p$ for polysacharides, 
		\item $B$ for \textit{Bacteroides thetaiotaomicron}, 
		\item $a$ for acetate, 
		\item $h$ for carbon dioxide and hydrogen,
		\item $M$ for \textit{ Methanobrevibacter
			smithii},
		\item $E$ for \textit{Eubacterium rectale}.
	\end{itemize}
	
	In \cite{Adrian_Ayati_Mangalam_2023} the dynamics are represented by a system of ODEs:
	\begin{subequations}\label{eq:ODE Bruce}
		\begin{align} 
			\frac{dB}{dt} & = \beta_B \Psi_{\gamma_p}(p)B -qB ,\\
			\frac{dE}{dt} &= \{ \beta_{E_1} \Psi_{\gamma_a}(a) + \beta_{E_2}(1 - \Psi_{\gamma_B}(B))\Psi_{\gamma_p}(p)\}E -qE, \\
			\frac{dM}{dt} &= \{ \beta_{M_1}\Psi_{\gamma_a}(a) + \beta_{M_2}\Psi_{\gamma_h}(h)\} M -qM, \\
			\frac{dp}{dt} &= \beta_pq(cos(t) +1)^3 -qp - \mu_{p,B}\Psi_{\gamma_p}(p)B \nonumber \\
			& \qquad - \mu_{p,E}(1 - \Psi_{\gamma_B}(B))\Psi_{\gamma_p}(p)E, \\
			\frac{da}{dt} &= \beta_a \Psi_{\gamma_p}(p)B -qa - \mu_{a,E} \Psi_{\gamma_a}(a)E - \mu_{a,M} \Psi_{\gamma_a}(a)M, \\
			\frac{dh}{dt} &= \beta_{h_1}\Psi_{\gamma_a}(a)E + \beta_{h_2}\Psi_{\gamma_p}(p)B + \beta_{h_3}(1 -\Psi_{\gamma_B}(B))\Psi_{\gamma_p}(p)E \nonumber \\ & \qquad -qh - \mu_{h,M} \Psi_{\gamma_h}(h)M. 
		\end{align} 
	\end{subequations}

	For further details and an explanation of the rest of the terms, we refer the reader to \cite{Adrian_Ayati_Mangalam_2023}.  We apply Deficiency Theory to begin to understand when this system exhibits multistationarity. First we need to translate the network's Monod kinetics into mass-action kinetics. As such we will need to make two simplifying assumptions:	
	\begin{enumerate}
		\item when $\Psi_{\gamma_x}$ is the only instance of a Monod function in a term in an equation, we assume saturation and set $\Psi_{\gamma_x}(x) = x$;   
		\item when a term in an equation includes the product $(1-\Psi_{\gamma_x}(x))(\Psi_{\gamma_y}(y))$, we set $\Psi_{\gamma_x}(x) = 1$ to eliminate that term in the equation.
	\end{enumerate}
The graph of the reaction network resulting from these simplifications is
	
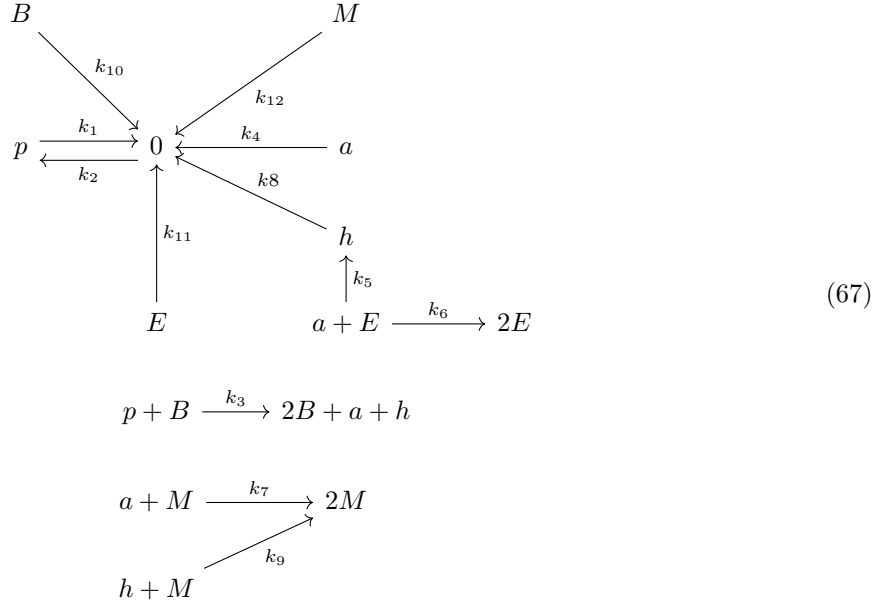
\begin{equation}
		\begin{tikzcd} \label{RN:project3 graph full network}
			B && M \\
			\\
			p & 0 & a \\
			&& h \\
			& E & {a+E} & 2E \\
			& {p+B} & {2B+a+h} \\
			& {a+M} & 2M \\
			& {h+M}
			\arrow["{k_1}", shift left, from=3-1, to=3-2]
			\arrow["{k_{10}}", from=1-1, to=3-2]
			\arrow["{k_{11}}"', from=5-2, to=3-2]
			\arrow["k8"', from=4-3, to=3-2]
			\arrow["{k_5}"', from=5-3, to=4-3]
			\arrow["{k_6}", from=5-3, to=5-4]
			\arrow["{k_2}", shift left=2, from=3-2, to=3-1]
			\arrow["{k_3}", from=6-2, to=6-3]
			\arrow["{k_9}"', from=8-2, to=7-3]
			\arrow["{k_{12}}", from=1-3, to=3-2]
			\arrow["{k_4}"', from=3-3, to=3-2]
			\arrow["{k_7}", from=7-2, to=7-3]
		\end{tikzcd}
	\end{equation}		
This treatment allows us to reduce the original ODE system in Equations \eqref{eq:ODE Bruce} to a system of polynomial equations,
	\begin{subequations}
		\begin{align} \label{eq:ODEs simplified project 3 RN}
			\frac{dp}{dt} &= k_2-k_1p-k_3pB, \\
			\frac{da}{dt} &= -k_4a +k_3pB -k_5aE +k_6aE -k_7aM, \\
			\frac{dh}{dt} &= -k_8h+k_5aE+k_3pB-k_9hM, \\
			\frac{dB}{dt} &=-k_{10}B +k_3pB, \\
			\frac{dE}{dt} &= -k_{11}E -k_5aE +k_6aE, \\
			\frac{dM}{dt} &= -k_{12}M +k_7aM +k_9hM.
		\end{align}
	\end{subequations}
For further simplification, we can set the rate constants to unity, \emph{e.g.,} $q=\mu_{x,y}=\beta_{x_i}=1$. These simplifying assumptions give us the system
	\begin{subequations}
		\begin{align}	
			\frac{dp}{dt} &= 1 -p - pB, \\
			\frac{da}{dt} &= pB -a - aE - aM, \label{ODE:project3 da pre-parameters}\\
			\frac{dh}{dt} &= aE + pB  -h - hM,  \label{ODE:project3 dh pre-parameters}  \\
			\frac{dB}{dt} & = pB -B, \\
			\frac{dE}{dt} &=  aE -E, \label{ODE:project3 dE pre-parameters}\\
			\frac{dM}{dt} &= (a + h)M -M. 
		\end{align}
	\end{subequations}
Putting aside reaction rates for the moment, we can now piece together how the species and complexes of the graph of the reaction network should be combined and where they should appear as reactants or products in the reactions. We note that the monomial $aE$ appears in Equations \eqref{ODE:project3 dE pre-parameters}, \eqref{ODE:project3 da pre-parameters}, and \eqref{ODE:project3 dh pre-parameters}.   We conclude that $a +E$ is a complex for the mass-action network and is the reactant complex to produce $h$ in a reaction. We also note that it is the reactant complex for a reaction that produces $2E$ as a product complex since it is positive in both Equations \eqref{ODE:project3 dE pre-parameters} and \eqref{ODE:project3 dh pre-parameters}. Since it is negative in Equation \eqref{ODE:project3 da pre-parameters} it never produces the species $a$ in any reaction. By systematically doing this type of analysis, we can produce the mass-action reaction network, where the last task is to assign rate parameters for each reaction.  Since we have 6 species and 12 reactions, our stoichiometric matrix $N \in \mathbb{R}^{6 \times 12}$ is 	
	\begin{equation} \label{computation:og network prj3:N}
		N = 
		\left[
		\begin{array}{*{12}c}
			-1 & 1 & -1 & 0 & 0 & 0 & 0 & 0 & 0 & 0 & 0 & 0 \\
			0 & 0 & 1 & -1 & -1 & -1 & -1 & 0 & 0 & 0 & 0 & 0 \\
			0 & 0 & 1 & 0 & 1 & 0 & 0 & -1 & -1 & 0 & 0 & 0 \\
			0 & 0 & 1 & 0 & 0 & 0 & 0 & 0 & 0 & -1 & 0 & 0 \\
			0 & 0 & 0 & 0 & -1 & 1 & 0 & 0 & 0 & 0 & -1 & 0 \\
			0 & 0 & 0 & 0 & 0 & 0 & 1 & 0 & 1 & 0 & 0 & -1
		\end{array}
		\right].
	\end{equation}	
	We immediately notice that because there is an outflow reaction ($X_i \to 0$) for each species in the network, we have a corresponding column vector in $N$ with $-1$ in the position of species $X_i$, and $0$ everywhere else. This forms a basis for $\mathbb{R}^6$, and thus $N$ has full rank. Furthermore, the corank of our network is zero, {\em i.e.,} $\rank(W) = 0$, and thus there are no conservation relations in our network. A consequence of this is that our network is not conservative and thus we cannot use Algorithm \ref{def:algorithm} as was done in Section \ref{sec:algorithm}. 
	
We can still glean some information on the mass-action reaction network in Equations \ref{RN:project3 graph full network} with Deficiency Theory. Using the chemical reaction network toolbox \cite{CRNT_toolbox}, we can determine if the network has the capacity to allow multiple steady states, \emph{i.e.,} there exists some assignment of rate parameters for the network in which there exists at least two distinct positive steady states. We can also study how the systematic removal of different flow reactions affects the capacity for multistationarity. We can thus determine which flow reactions are necessary for monostationarity for the system and furthermore when multistationarity is possible.
	
We first notice that the only species for which we have an inflow reaction, {\em i.e.,} $0 \to X_i$, is the species $p$. For every other species we only have outflow reactions, {\em i.e.,} $X_i \to 0$.  We use variable names in the first column as shorthand for the subnetwork obtained by ignoring the flow reactions involving those species, \emph{e.g.,} entry $a,B$ corresponds to the network defined in Equation \eqref{RN:project3 graph full network}, where we removed outflow reactions $a \to 0$ and $B \to 0$. Since we have both inflow and outflow reactions for $p$, we denote $p+$ as the outflow reaction $p \to 0$ and $p-$ as the inflow reaction $0 \to p$. This subnetwork has the graph
	
	
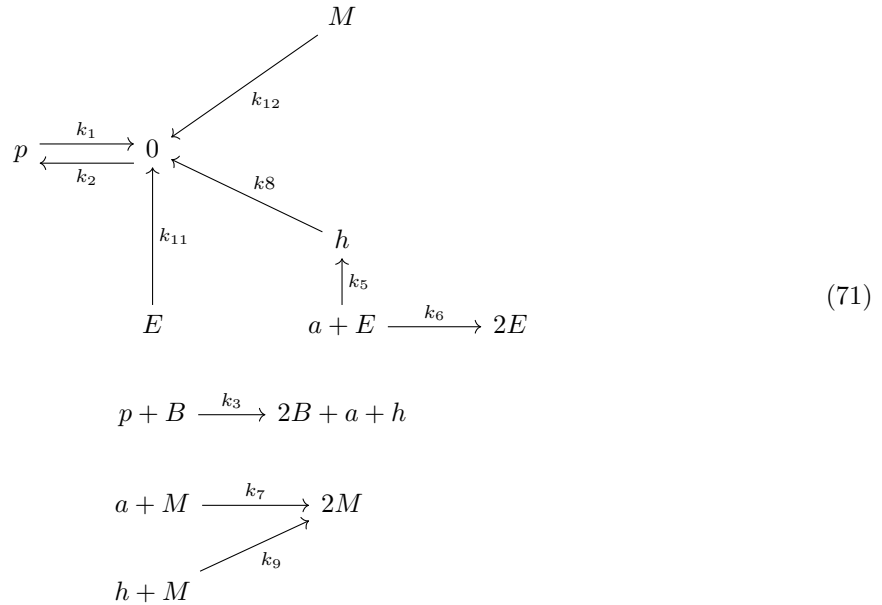
\begin{equation} \label{RN:bacteria a,B removed}
		\begin{tikzcd}
			&& M \\
			\\
			p & 0 \\
			&& h \\
			& E & {a+E} & 2E \\
			& {p+B} & {2B+a+h} \\
			& {a+M} & 2M \\
			& {h+M}
			\arrow["{k_{12}}", from=1-3, to=3-2]
			\arrow["{k_1}", shift left, from=3-1, to=3-2]
			\arrow["{k_2}", shift left=2, from=3-2, to=3-1]
			\arrow["k8"', from=4-3, to=3-2]
			\arrow["{k_{11}}"', from=5-2, to=3-2]
			\arrow["{k_5}"', from=5-3, to=4-3]
			\arrow["{k_6}", from=5-3, to=5-4]
			\arrow["{k_3}", from=6-2, to=6-3]
			\arrow["{k_7}", from=7-2, to=7-3]
			\arrow["{k_9}"', from=8-2, to=7-3]
		\end{tikzcd}
	\end{equation}
	
Recall that since since the network in Equation \eqref{RN:project3 graph full network} had a flow reaction for each species, the network is not conservative. We checked, using the chemical reaction network toolbox \cite{CRNT_toolbox}, if any subnetwork generated by excluding any combination of flow reactions would result in a conservative network.  None did.
	
We use three distinct categories for the number of steady states: 
	\begin{description}
		\item[Monostationary.] For some assignment of rate constants, the system admits at most one positive steady state.  This does not mean that any assignment of rate constants yields a monostationary reaction network and furthermore not every stoichiometric compatability class will contain a positive steady state; this just precludes the scenario where an assignment of rate constants gives rise to multiple steady states in a single stoichiometric compatability class.
		\item[Possibly Multistationary] It is possible that the network has a rate constant assignment that gives rise to multiple positive steady states in a stoichiometric compatability class. Currently the CRNT Toolbox \cite{CRNT_toolbox} cannot preclude multistationarity for these reaction networks. A different analyis is needed to establish multistationarity.
		\item[No positive steady states.] Regardless of rate constant values, none of the stoichiometric compatability classes admit a positive steady state.
	\end{description} 

Since we are considering 7 flow reactions, this corresponds to considering $2^7 = 128$ reaction networks, including the full network (the graph in Equation \eqref{RN:project3 graph full network}). We summarize the results from the CRNT Toolbox \cite{CRNT_toolbox} for these cases in Table \ref{table:chemostat stationarity}.

\begin{table}[]
\begin{tabular}{|l|l|}
\hline
Steady-state Category & Outflow Eliminated from Model \\
\hline
Monostationary (6) & 0, $a$, $p+$, $ah$, $ap+$, $ahp+$ \\
\hline
Possibly Multistationary (10) & $h$, $E$, $aE$, $hp+$, $hE$, $p+E$, $ahE$, $ap+E$, $hp+E$, $ahp+E$ \\
\hline
No Positive Steady States (112) & All other formulations \\
\hline 
\end{tabular}
\caption{Steady-state information on models excluding certain outflow reactions. Here the entry labeled $0$ corresponds to the full model in Equation \eqref{RN:project3 graph full network}. Note that $aB$ would fall under having no positive steady states, and the reaction network would correspond to Equation \eqref{RN:bacteria a,B removed}.}
\label{table:chemostat stationarity}
\end{table}

From Table \ref{table:chemostat stationarity}, we can make a few observations regarding the removal of outflow reactions involving certain species. Removal of outflow reactions involving species $p-, B, M$ totally destroy the systems ability to obtain a single positive steady state. We can also note that removal of the outflow reaction involving species $E$ always gives rise to a subnetwork with the capacity for multistationarity. This highlights the influence that the rate constants for the outflow reactions involving $p-,B,M,E$ have on the number of steady states of this model.
	
	\section{Conclusions}
	\label{sec:concl}
	The tools presented in this paper come primarily from abstract algebra and its relation to chemical reaction network theory.  Specifically, we have taken advantage of the fact that the system of polynomials describing steady states of mass-action systems are Gr\"obner bases for the ideals of all such equations, and used that fact towards qualitative analysis, in our case determining the existence and number of positive steady states. Our goal has been to move towards understanding the behavior of reaction networks without needing to specify reaction rates.  The work in this paper left an open question, a conjecture of sorts mentioned in Remark \ref{remark:conjecture}.
	
	The general approach we took was to see what could be had by decomposing networks into subnetworks.  Building understanding by rejoining the subnetworks remains a future direction.  
	
	The qualitative theory of ordinary differential equations has had a tremendous impact over the last decades, with many models now beginning life as qualitative representations.  Perhaps most famously is the parameter $R_0$ from epidemiology that arose from such an approach.  Creating a qualitative theory based on abstract algebra is a relatively unexplored direction, but in our view holds promise.

	\paragraph{Acknowledgements} 
	The authors thank Prof. Colleen Mitchell and Prof. Zahra Aminzare for their insights, suggestions, and conversations. JMS thanks Prof. Elisenda Feliu and Prof. Alicia Dickenstein for their wonderful summer school through MSRI in Leipzig, and Skye Dore-Hall and Jos\'{e} Lozano for their contributions to the examples in Equations \eqref{RN:1 site phosphorylation dephosphorylation modified} and \eqref{eg:2 site graph}.
	
	\bibliography{biblio}

\end{document}